\documentclass[aps,prd, amsmath, preprint, preprintnumbers, showpacs, floatfix]{revtex4-1}
\usepackage[dvips]{graphicx}
\usepackage{amssymb}
\usepackage{subfigure}
\usepackage{multirow}

\usepackage{color}

\begin{document}
%\preprint{}
\preprint{CU-TP-1201}

\title{Light Quark Mass Reweighting}

\newcommand\cu{Physics Department, Columbia University, New York, NY 10027, USA}
\newcommand\bnl{Brookhaven National Laboratory, Upton, NY 11973, USA}

\author{Qi Liu}\affiliation{\cu}
\author{Norman H. Christ}\affiliation{\cu}
\author{Chulwoo Jung}\affiliation{\bnl}

\collaboration{RBC and UKQCD Collaborations}

\date{May 31, 2012}

\begin{abstract}
We present a systematic study of the effectiveness of light quark mass reweighting.  This method allows a single lattice QCD ensemble, generated with a specific value of the dynamical light quark mass, to be used to determine results for other, nearby light dynamical quark masses.  We study two gauge field ensembles generated with 2+1 flavors of dynamical domain wall fermions with light quark masses  $m_l=0.02$ ($m_\pi=620$ MeV) and $m_l=0.01$ ($m_\pi=420$ MeV).   We reweight each ensemble to determine results which could be computed directly from the other and check the consistency of the reweighted results with the direct results.   The large difference between the 0.02 and 0.01 light quark masses suggests that this is an aggressive application of reweighting as can be seen from fluctuations in the magnitude of the reweighting factor by four orders of magnitude.  Never-the-less, a comparison of the reweighed topological charge, average plaquette, residual mass, pion mass, pion decay constant, and scalar correlator between these two ensembles shows agreement well described by the statistical errors.  The issues of the effective number of configurations and finite sample size bias are discussed.  An examination of the topological charge distribution implies that it is more favorable to reweight from heavier mass to lighter quark mass.
\end{abstract}

\maketitle

\section{Introduction}
Generating ensembles of gauge field configurations with dynamical quarks is usually the most expensive part of a lattice QCD calculation. Because of the renormalization of the quark mass, we are usually not able to identify in advance the input bare quark mass that will correspond to a target renormalized mass.  Quark mass reweighting is a powerful technique that allows us to fine tune the sea quark masses to their physical or other desired values after an ensemble has been generated, avoiding the computationally expensive generation of new ensembles.  As is explained below, a reweighted lattice quantity is computed by averaging the product of that quantity and a reweighting factor over the given ensemble.  The reweighting factor is chosen to reproduce the effects of the change in the action that would result from changing from the simulated to the reweighted quark mass.

Reweighting of the strange quark mass has been widely applied to accomplish such a fine tuning of the physical strange quark mass~\cite{Aoki:2010pe,Aoki:2010dy,Christ:2010dd} or to study dependence on the dynamical strange quark mass~\cite{Ohki:2009mt}.  Reweighting of the light quark mass has been used less frequently, presumably because the light quark masses used in most current lattice calculations are sufficiently heavy that reweighting methods may not be able to bridge the gap between the simulated and physical masses.  However,  Aoki {\it et al.}~\cite{Aoki:2009ix}  have applied light quark mass reweighting to obtain results at the physical light quark mass.  In the future,  we expect there will be an increasing use of mass reweighting to adjust the light quark mass to its physical value.  Another important application is the use of light quark mass reweighting to avoid difficulties caused by a small quark mass such as long autocorrelation times or a loss of numerical stability at small mass.  For example in the case Wilson fermions, one intentionally performs a simulation at larger light quark mass and then reweights to the physical mass~\cite{Hasenfratz:2008fg, Luscher:2008tw}. This can also potentially reduce the computational cost. 

As an important and widely used technique in lattice QCD, quark mass reweighting has been carefully studied and examined, for example, using Wilson~\cite{Hasenfratz:2008fg} or overlap~\cite{DeGrand:2008ps} fermions.  However, to better understand how mass reweighting performs, we believe that addition study of the resulting statistical fluctuations and a systematic comparison of a reweighted ensemble with one directly generated without reweighting are needed. As has been emphasized in earlier work such as Refs.~\cite{Hasenfratz:2008fg,DeGrand:2008ps, Luscher:2008tw}, a small overlap in configuration space between the simulated ensemble and the target ensemble can lead to large fluctuations in the reweighting factors and unreliable results because of the lack of sufficient statistics. There are two issues. First, for a fixed sample size N, the reweighting factor is determined from what may be a biased estimator.  (We will discuss this in detail in section IV). If the reweighting factors fluctuate too much, this bias may lead to a large systematic error. Second, without a sufficiently large number of configurations with non-negligible weight, the error estimate may not be reliable, or even when reliable, the estimated may be too large for the result to be useful.

In this paper, we provide a systematic check of the reweighting technique by using two ensembles generated with the Iwasaki gauge action and the Domain Wall Fermion (DWF) action for the quarks. In order to see the limitations of reweighting, we consider a case were the method is used to achieve a large change in fermion mass.   Specifically we reweight the light input quark mass from 0.02 ($m_\pi=620$ MeV) to 0.01($m_\pi=420$ MeV) and then check the reweighted results for many important physical quantities against those computed directly on the dynamical $m_l=0.01$ ensemble. We also do this in the reverse direction, reweighting the $m_l=0.01$ ensemble to 0.02.  With the ensemble size up to 1000 configurations (a total of 10,000 molecular dynamics time units), we find that even for such a large range of quark mass reweighting, the reweighted results and those obtained directly on the dynamically generated ensemble are consistent, demonstrating the success of the reweighting technique. 

We study the behavior of reweighting on both short and long distance quantities.  The paper is organized as follows:  We begin in Sec.~\ref{sec:method} with a description of the reweighting method, a detailed study of the reweighting factors and an analysis of the effective number of configuration that results after reweighting.  We then discuss the reweighting of the topological charge and susceptibility in Sec.~\ref{sec:topology}.   In Sec.~\ref{sec:plaquette}, we discuss the reweighting of a short distance physical quantity, the average plaquette, and the problem of reweighting bias.  A second short distance quantity which shows significant dynamical quark mass dependence, the residual mass, is examined  in Sec.~\ref{sec:pion_fpi_mres} as well as two important physical observables: the pion mass and decay constant,.  Finally the effect of reweighting on the scalar correlation function is presented in Sec.~\ref{sec:correlator} and in Sec.~\ref{sec:summary} we further discuss and summarize our results.

\section{Light quark mass reweighting method}
\label{sec:method}
We follow closely the strategy presented in Ref~\cite{Hasenfratz:2008fg}.  Reweighting an ensemble of configurations from the light sea quark mass $m_1$ to $m_2$ requires the evaluation of a reweighting factor $w(U;m_1,m_2)$ for each configuration on which measurements are to be performed.  We can see that the expectation value for any operator $O$ on an ensemble with light sea quark mass $m_2$ can be calculated from an ensemble generated with sea quark mass $m_1$ as follows.  One begins with the usual expression for the expectation value of the operator $O$ computed for the case of a sea quark mass $m_2$: 
\begin{eqnarray}
  \left<O\right>_2 & = & \frac{1}{Z_2}\int DU\, O\, e^{-S_g} \frac{\det\{D^\dagger (m_2)D(m_2)\}}{\det\{D^\dagger (1.0)D(1.0)\}}\frac{\sqrt{\det\{D^\dagger (m_s)D(m_s)\}}}{\sqrt{\det\{D^\dagger (1.0)D(1.0)\}}}.
\label{eq:rwo_0}
\end{eqnarray}
Here $D(m)$ represents the DWF Dirac operator for a fermion of mass $m$, $m_s$ is the strange quark mass and the determinant factors with mass $m=1.0$ are the standard Pauli-Villars contributions which appear in the DWF formulation.  The quantity $Z_2$ appearing in the denominator is the partition function for the sea quark mass $m_2$: simply the integral in Eq.~\eqref{eq:rwo_0} with the factor in the numerator representing the operator $O$ removed.  

Next one multiplies and divides by factors of $Z_1$ and $\det\{D^\dagger (U,m_1)D(U,m_1)\}$ obtaining
\begin{eqnarray}
  \left<O\right>_2 && \\ 
& & \hskip -0.2 in = \frac{Z_1}{Z_2} \cdot \frac{1}{Z_1}\int DU \left[O\,\frac{\det\{D^\dagger (m_2)D(m_2)\}}{\det\{D^\dagger (m_1)D(m_1)\}}\right] e^{-S_g} \frac{\det\{D^\dagger (m_1)D(m_1)\}}{\det\{D^\dagger (1.0)D(1.0)\}}\frac{\sqrt{\det\{D^\dagger (m_s)D(m_s)\}}}{\sqrt{\det\{D^\dagger (1.0)D(1.0)\}}} \nonumber \\
& & \hskip -0.2 in =  \frac{\left<O \, w(m_1,m_2)\right>_1}{\left<w(m_1,m_2)\right>_1}.  \label{eq:rwo}
\end{eqnarray}
Here we have introduced the reweighting  factor $w(U;m_1,m_2)$ for each configuration $U$:
\begin{eqnarray}
w(U;m_1,m_2) & = & \frac{\det\{D^\dagger (U,m_2)D(U,m_2)\}}{\det\{D^\dagger (U,m_1)D(U,m_1)\}}
\end{eqnarray}
and performed a similar manipulation to write the ratio $Z_1/Z_2$ as $\left<w(m_1,m_2)\right>_1$.  
In the above formulae, we have used the notation $D(U,m)$ to make explicit the dependence of the DWF Dirac operator on the gauge field variable $U$.   The reweighting factor $w(U;m_1,m_2)$ can be determined from the stochastic average:
\begin{eqnarray}
w(U;m_1,m_2) & = & \frac{\int D\xi e^{-\{\xi^\dagger D(m_1)^\dagger D(m_2)^{\dagger -1}D(m_2)^{-1}D(m_1) \xi-\xi^\dagger\xi\}}e^{-\xi^\dagger\xi}}{\int D\xi e^{-\xi^\dagger\xi}} \\
& = & \left<e^{- \{\xi^\dagger D(m_1)^\dagger D(m_2)^{\dagger -1}D(m_2)^{-1}D(m_1)\xi - \xi^\dagger\xi\}}\right>_\xi  \label{eq:rwexpect},
\end{eqnarray}
where the average appearing in Eq.~\eqref{eq:rwexpect} is to be performed over the stochastic variable $\xi$ drawn from the random Gaussian distribution $\exp\{-\xi^\dagger\xi\}$.  

We have purposely arranged the product of the four $D(U,m)$ operators in the exponent of Eq.~\eqref{eq:rwexpect} in a form that will minimize the cost of the needed Conjugate Gradient (CG) inversion. Let us call the exponent of that equation $f(U; m_1,m_2; \xi)$ so that $w(U;m1,m2) = \left<e^{-f(U;m_1,m_2; \xi)}\right>_\xi$. Then $f(U;m_1,m_2; \xi)=\eta^\dagger\eta-\xi^\dagger\xi$, where 
\begin{equation}
\eta=D(m_2)^{-1}D(m_1)\xi = [D(m_2)^\dagger D(m_2)]^{-1} D(m_2)^\dagger D(m_1)\xi. \label{eq:eta}
\end{equation}
If $m_2$ is close to $m_1$, then the solution for $\eta$ will usually converge faster than will the solution for a combination such as $[D(m_2)^\dagger D(m_2)]^{-1}D(m_1) \xi$, which would have resulted from a different arrangement than that used above. 

It was pointed out in Ref~\cite{Hasenfratz:2008fg} that the average of stochastic sources and the average of gauge fields can be performed in any order, so we can perform the average needed in  Eq.~\eqref{eq:rwexpect} using any number of random hits $N_{\rm hit}$.  By doing so, we introduce extra noise into our system, but no systematic error.  To reduce the noise following the idea of determinant factorization proposed by Hasenbush~\cite{Hasenbusch:2001ne,Hasenfratz:2008fg}, we calculate the reweighting factor by carrying out $N_{\rm step}$ steps.  For each step we change the mass by $\Delta m = (m_2-m_1)/N_{\rm step}$. Following this approach, the reweight factor for a given configuration is then calculated from the product
\begin{eqnarray}
w(U; m1,m2) &=& \prod_{s=1}^{N_{\rm step}} w(U,m_1+(i-1)\Delta m,  m_1+i\Delta m)  \\
&=& \prod_{s=1}^{N_{\rm step}}\left\{\frac{1}{N_{\rm hit}}\sum_{i=1}^{N_{\rm hit}} e^{- f(U; m_1+(i-1)\Delta m, m_1+ i \Delta m; \xi_{i,s}) }\right\}. \label{eq:wcal}
\end{eqnarray}
where for each step, $N_{\rm hit}$ hits are performed by summing over the $N_{\rm hit}$ Gaussian variables $\{\xi_{i,s}\}_{1 \le i \le N_{\rm hit}}$ for each of the steps $s$, $1 \le s \le N_{\rm step}$.

Our goal is to calculate the reweighting factor with sufficient accuracy that the uncertainty introduced by the finite number of the random hits is smaller than the fluctuation of $w(U; m1,m2)$ among the underlying gauge configurations. The total number of DWF Dirac inversions required by Eq.~\eqref{eq:wcal} is $N_t=N_{\rm hit} N_{\rm step}$. By increasing $N_{\rm step}$, we reduce the fluctuation among the independent steps. Averaging over an additional $N_{\rm hit}$ hits gives a better estimate of the average. From our experiments with a fixed number of total inversions, we find that smaller errors can be achieved from more steps than more hits.  These results confirm similar conclusions reached by the RBC and UKQCD collaborations when reweighting in the strange quark mass.

This observation can be verified by the following explicit calculation.  Consider a single gauge configuration and evaluate the reweighting factor $w$ as a product of the reweighting factors $w_s$, $1 \le s \le N_{\rm step}$ corresponding to $N_{\rm step}$ steps.  In our stochastic evaluation the factor $w_{s}$ corresponding to the step $s$ is given by a function $w_{s}(\xi)$ of a set of random variables $\xi$  where a different stochastic vector $\xi$ will be used for each step.  Define the quantity $\sigma_s$ from the ratio of averages:
\begin{equation}
\frac{\left\langle w_{s}(\xi)^2 \right\rangle}{\langle w_{s}(\xi)\rangle^2} = e^{\sigma^2_s/N_{\rm step}^2}
\end{equation}
where the average is taken over the stochastic vector $\xi$ in the fixed gauge background.  The factor $1/N_{\rm step}^2$ has been extracted from the exponent on the right hand side because the exponent in Eq.~\eqref{eq:rwexpect} will be reduced in size by $1/N_{\rm step}$ when a fixed mass interval is divided into $N_{\rm step}$ segments.  Since the reweighting factor $w$ is linear in each of the independent stochastic factors $w_{s}(\xi_s)$, it is a straight forward exercise to express the fluctuations in $w$ in terms of the quantities $\sigma_s$.  If $w$ is obtained from a single hit:
\begin{eqnarray}
(\delta w)^2 &=& \left\langle(w-\langle w\rangle)^2\right\rangle \label{eq:stoch_calc_1}\\
                   &=& \left\langle w^2\right\rangle - \langle w\rangle^2 \\
                   &=& \prod_{s=1}^{N_{\rm step}} \left\langle w_s^2\right\rangle -  \prod_{s=1}^{N_{\rm step}} \langle w_s\rangle^2 \\
                   &=&  \langle w\rangle^2 \left( \prod_{s=1}^{N_{\rm step}} e^{\sigma_s^2/N_{\rm step}^2} -1  \right) \\
                   &=&  \langle w\rangle^2 \left( e^{\sigma^2/N_{\rm step}} -1 \right), \label{eq:stoch_calc_2}
\end{eqnarray}
where in the final line we have made the simplifying assumption that $\sigma_s$ is independent of $s$.  If this process is repeated $N_{\rm hit}$ times and the results averaged ($N_{\rm hit}$ hits) then the resulting error from the stochastic sampling becomes
\begin{eqnarray}
(\delta w) =  \langle w\rangle\sqrt{ \frac{e^{\sigma^2/N_{\rm step}} -1}{N_{\rm hit}}},
\label{eq:w_fluct_1}
\end{eqnarray}
A smaller error results if the $N_{\rm hit}$ hits are averaged at each step, as is indicated in Eq.~\eqref{eq:wcal}.  For this case a series of steps similar to those in Eqs.~\eqref{eq:stoch_calc_1}-\eqref{eq:stoch_calc_2} gives:
\begin{equation}
\delta w = \langle w\rangle \left\{\left(1+\frac{1}{N_{\rm hit}}(e^\frac{\sigma^2}{N_{\rm step}^2} -1)\right)^{N_{\rm step}} - 1\right\}^{1/2}.
\label{eq:w_fluct_2}
\end{equation}
where we write the simpler expression that results if we continue to use uniform steps and assume that the corresponding quantities $\sigma_s$ are independent of $s$.

While the fluctuations given in Eq.~\eqref{eq:w_fluct_2} are smaller than those in  Eq.~\eqref{eq:w_fluct_1}, both expressions are decreasing functions of $N_{\rm step}$ for a fixed total number $N_t$ of Dirac inversions, making it favorable to use larger $N_{\rm step}$ instead of larger $N_{\rm hit}$.  When $N_{\rm step}$ is sufficient large, both of the above formulae become approximately $w\sigma/\sqrt{N_t}$.  In this limit it is unimportant how $N_t$ is divided between steps and hits.   A second advantage of introducing a number of Hasenbusch steps is that we can then reweight to many different intermediate mass values in a single reweighting calculation. A third advantage can be seen from Eq.~\eqref{eq:eta}: splitting the ratio $m_2/m_1$ into more factors makes the mass difference appearing in each step smaller so that each inversion converges faster.  Thus, in all our reweighting calculations, we use only one hit, but many intermediate mass steps.

\begin{figure}[tbh]
  \begin{tabular}{ll}
	\includegraphics[width=80mm]{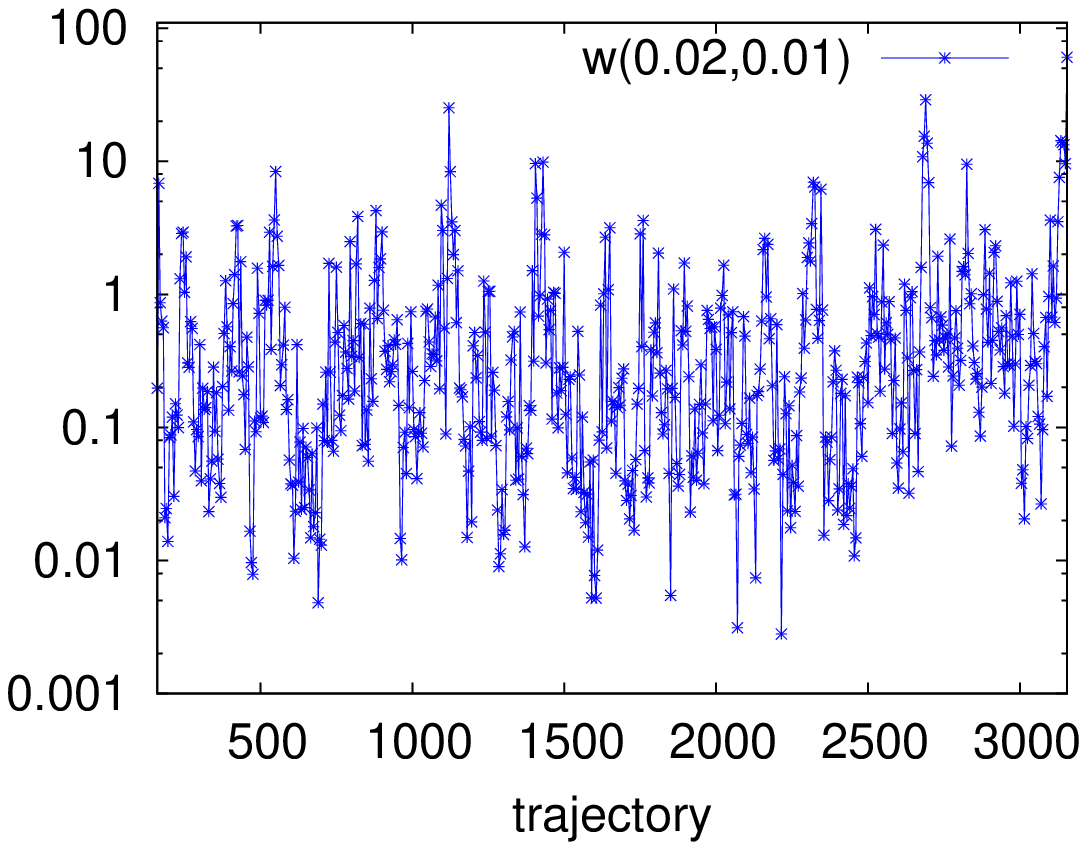} &
	\includegraphics[width=80mm]{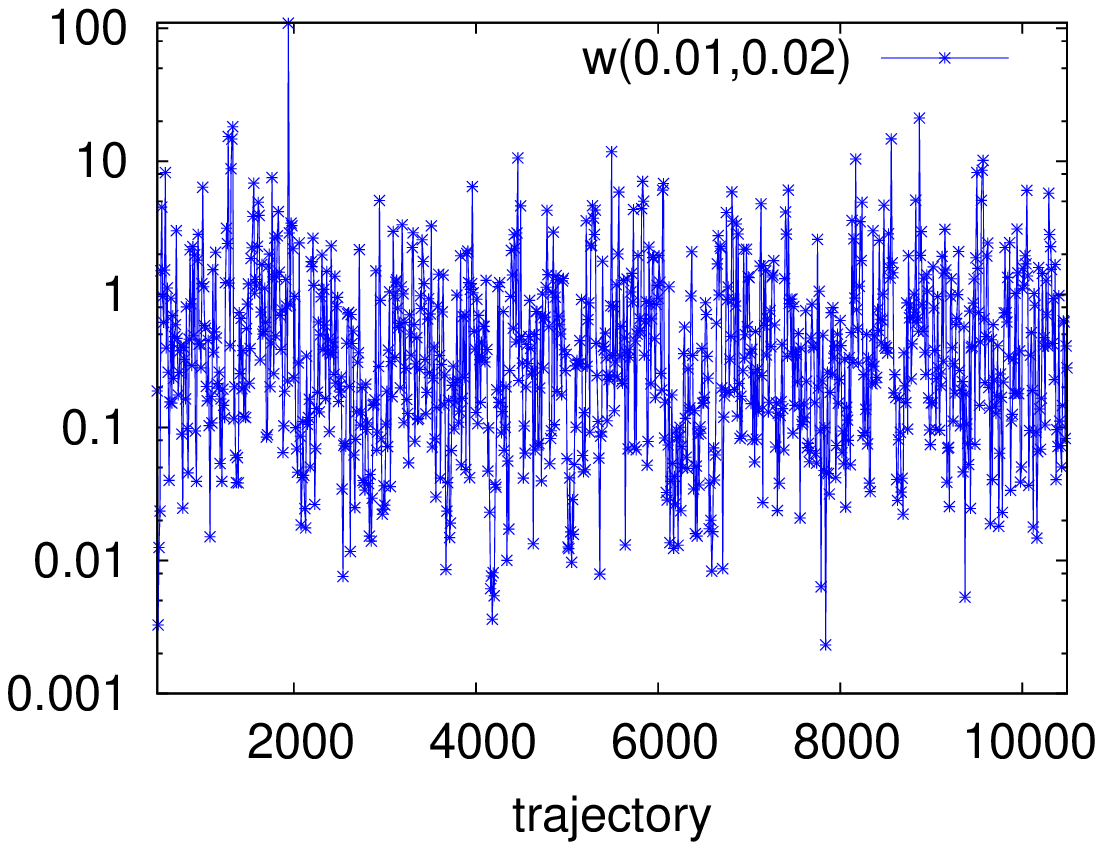} \\
  \end{tabular}
  \caption{Left: normalized reweighting factors computed from the $m_l=0.02$ configurations which can be used to reweight the light quark mass from $m_l=0.02 \to 0.01$. Right: normalized reweighting factors computed from the $m_l=0.01$ configurations which can be used to reweight from $m_l=0.01 \to 0.02$.  Here we plot the normalized reweighting factors $\hat w_i = N w_i/\sum_j w_j$ which become unity in the case of equal weights.}
\label{fig:rwfactor}
\end{figure}

We use two ensembles generated with 2+1 flavors of domain wall fermions, the Iwasaki gauge action with $\beta=2.13$, a $16^3\times32$ space-time volume and the extent $L_s=16$ in the fifth dimension.. The lattice spacing for this $\beta$ value has been determined to be $a^{-1}=1.73(3)$GeV~\cite{Allton:2008pn}. The configurations we use begin with  trajectory 160 and extend to trajectory 3155 sampled every 5 trajectories for the $m_l=0.02$ ensemble (600 configurations in total)  and extend from trajectory 500 to 10490, sampled every 10 trajectories, for the $m_l=0.01$ ensemble (1000 configurations in total). We reweight the $m_l=0.02$ ensemble to $m_l=0.01$, and $m_l=0.01$ to $m_l=0.02$ through 80 steps performing only one hit for each configuration. The calculated reweighting factors are shown in Fig.~\ref{fig:rwfactor}. 

Because of the large shift in mass that we are attempting to achieve with reweighting, the reweighting factors presented in Fig.~\ref{fig:rwfactor} vary in size by a few orders of magnitude. Such large fluctuations in the reweighting factors may cause concern that such a substantial reweighting is doomed to fail. However, the large fluctuations of these factors toward lower values are not important.  No mater how small they are, their only consequence is that the corresponding configurations make a negligible contribution in the reweighted ensemble. By simply counting the number of configurations that have reweighting factor larger than 1, we see a variation in size of 100 and 200 for the 0.02 and 0.01 ensembles respectively. Thus, the reweighting may be more successful than Fig.~\ref{fig:rwfactor} naively suggests. The efficiency of reweighting can be estimated in a more systematic way as follows.

%106 and 230

Suppose that we have a specific ensemble of N configurations.  The reweighted value for any physical quantity $O$ based on Eq.~\eqref{eq:rwo} is
\begin{equation}
\left<O_w\right>_N = \frac{\sum_{i=1}^N w_i O_i}{\sum_{j=1}^N w_j}.  \label{eq:OwN}
\end{equation} 
With the assumption that the reweighting factors $w_i$ and the measurements $O_i$ are weakly correlated, we can derive an approximate expression for the statistical fluctuations of the reweighted average $\left<O_w\right>_N$ around the average value $\left<\!\left<\left<O_w\right>_N\right>\!\right>$.  Here the brackets $\left<\!\left<\ldots\right>\!\right>$ indicate the result that would be
obtained when many ensembles of $N$ configurations are generated and the results of each such calculation averaged.  Of course, such an average will normally give the true value, {\it i.e.} the value in the limit of large $N$, except for quantities which are not simple averages where a bias may result if $N$ is insufficiently large.  If we keep only the dominant term we find (for a more detailed discussion and a derivation, see Appendix C of Ref.~\cite{Aoki:2010dy}):
\begin{equation}
  \left<\!\left<\Bigl(\left<O_w\right>_N -\left<\!\left< \left<O_w\right>_N\right>\!\right>\Bigr)^2\right>\!\right> \approx \delta O^2 \frac{\tau_{\rm corr}}{N}\frac{\overline{w^2}}{\overline{w}^2} \label{eq:rwerr}
\end{equation}
where the quantities appearing on the right-hand side can be approximately calculated from a single ensemble of $N$ configurations according to the equations:
\begin{eqnarray}
\delta  O^2 &  =  & \frac{1}{N}\sum_{i=1}^N (O_i - \left<O\right>_N)^2 \\
\overline{w} & = & \frac{1}{N}\sum_{i=1}^N w_i \\
\overline{w^2} & = & \frac{1}{N}\sum_{i=1}^N w_i^2 \\
\tau_{\rm corr} & = & \sum_{l=-L_{\rm max}}^{L_{\rm max}}C(l)W(l) \label{eq:autocorr}.
\end{eqnarray}
The quantities $C(l)$ and $W(l)$ appearing in Eq.~\eqref{eq:autocorr} are the autocorrelation functions for the measured quantity, $O_i$ and the the reweighting factors $w_i$ and can be estimated from a single ensemble using the formulae:
\begin{eqnarray}
  C(l) & = &\frac{1}{N-l}\sum_{i=1}^{N-l} \frac{(O_i - \left<O\right>_N)(O_{i+l} - \left<O\right>_N)}{\delta O^2} \label{eq:Cl} \\
  W(l) & = &\frac{1}{N-l} \sum_{i=1}^{N-l} \frac{w_i w_{i+l}}{\overline{w^2}}, \label{eq:Wl}
\end{eqnarray}
for the case $l\ge 0$.  When evaluating the autocorrelation time with a finite sample size N, we must impose a maximum length $L_{\rm max} \ll N$. Here, we use the smallest value of $l$ at which $C(l+1)$  becomes negative. 

If we interpret the error given in Eq.~\eqref{eq:rwerr} as the size of the fluctuations among our samples divided by the square root of an effective number of configurations, $\sqrt{\delta O^2/N_{\rm eff}}$, we can determine $N_{\rm eff}$ to be:
\begin{equation}
N_{\rm eff} = \frac{N}{\tau_{\rm corr}}\frac{\overline{w}^2}{\mbox{ $\overline{w^2}$ }}  \label{eq:Neff}
\end{equation}
In the case of no reweighting($w_i\equiv1$), this expression reverts to our normal expression for the effective number of configurations $N/\tau_{\rm corr}$. Even though different physical quantities have different autocorrelation times,  the reweighted auto correlation times $\tau_{\rm corr}$ in Eq.~\eqref{eq:autocorr} are usually very close to 1 for our data, because $C(l)$ and $W(l)$ are both small numbers except when l=0. Therefore, the estimated effective number of configurations can be calculated from the reweighting factors alone.  We find $N_{\rm eff}=48$ for the $m_l=0.02$ ensemble (reweighting to 0.01) and $N_{\rm eff}=63$ for the $m_l=0.01$ ensemble (reweighting to 0.02).  Clearly, the effective sample size decreases dramatically because of our large range of mass reweighting. However, with a sample size of roughly 50 configurations, we can expect that the reweighting still works well for many interesting physical quantities.

\begin{figure}[!h]
  \begin{tabular}{ll}
  \includegraphics[width=80mm]{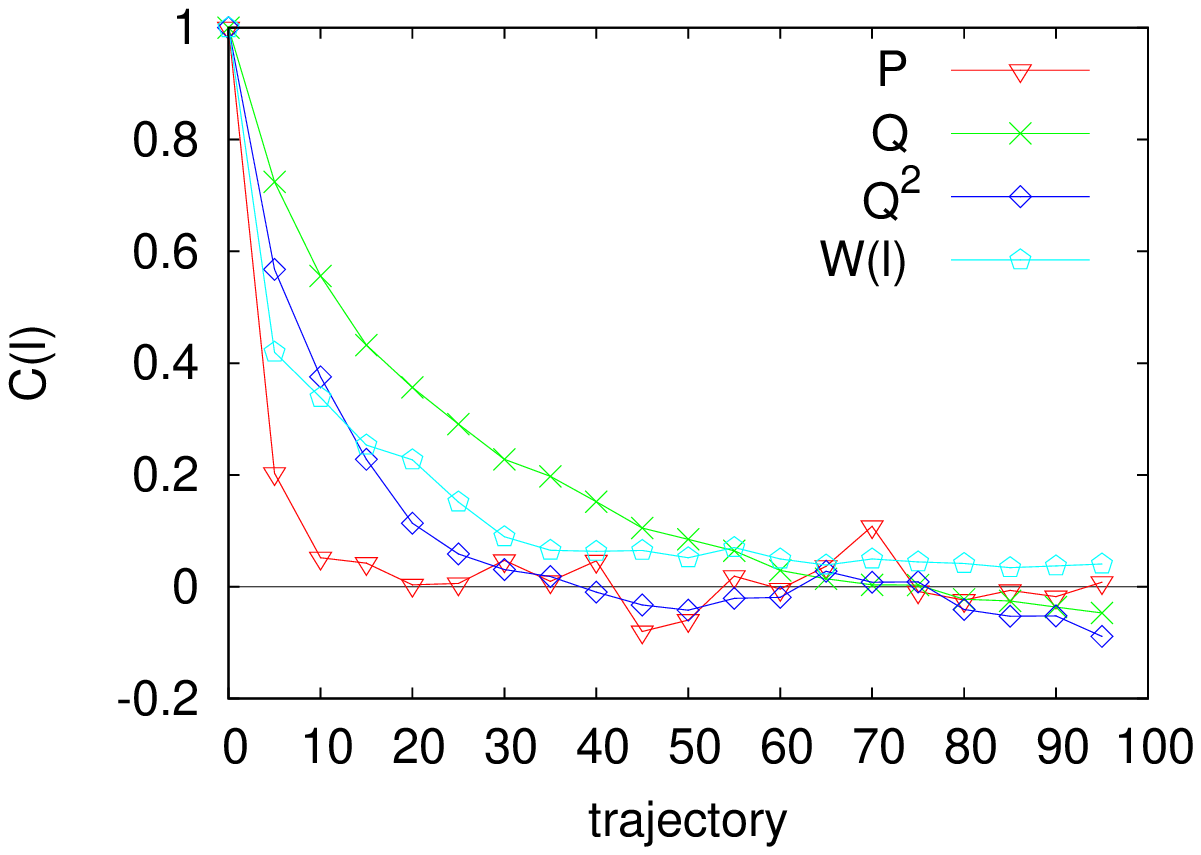} &
  \includegraphics[width=80mm]{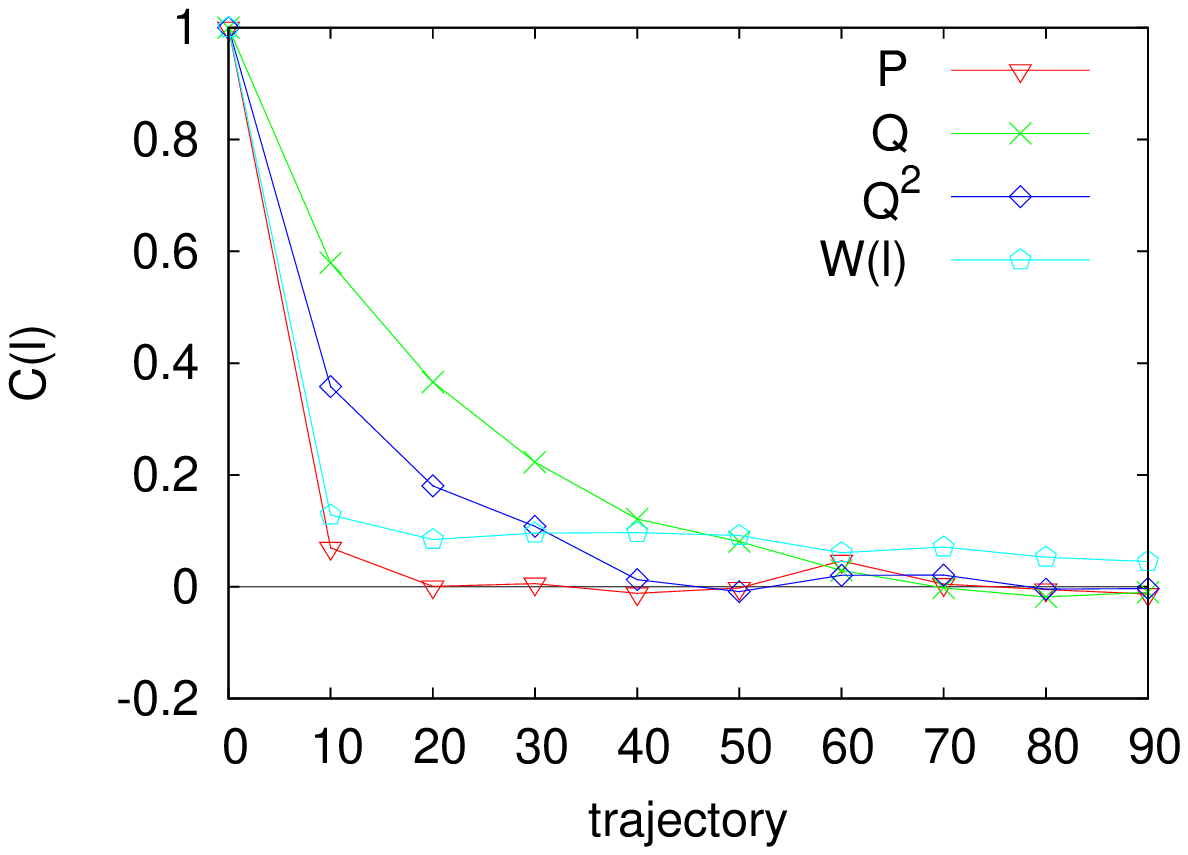} \\
  \end{tabular}
  \caption{The autocorrelation function $C(l)$ defined in Eq.~\eqref{eq:Cl} plotted versus $l$ for the average plaquette ($P$) , topological charge ($Q$) and $Q^2$.  The autocorrelation function W(l) for the reweighting factor defined in Eq.~\eqref{eq:Wl} is also shown.  The left panel shows results calculated from the $m_l=0.02$ ensemble while that on the right shows the results from the $m_l=0.01$ ensemble.}
  \label{fig:autocorr}
\end{figure}

To justify our assumption that $\tau_{\rm corr}\sim 1$, we calculated the autocorrelation function $C(l)$ for a few of quantities that we will study later.  Figure~\ref{fig:autocorr} shows the autocorrelation function $C(l)$ of the average plaquette(P), topological charge(Q) and susceptibility ($Q^2$) as well as the autocorrelation function $W(l)$ defined in Eq.~\eqref{eq:Wl}. The $W(l)$ function converges to the constant value $\overline{w}^2/\overline{w^2}$ for large separations. From these results the  integrated autocorrelation times $\tau_{\rm corr}$ defined by Eq.~\eqref{eq:autocorr} for the unreweighted and reweighted ensembles are calculated and summarized in Table~\ref{Tab:taucorr}. We can see that even for the topological charge which has the longest autocorrelation time, the reweighted integrated autocorrelation time is less than 2. Since most quantities have significantly shorter autocorrelation time compared to that of the topological charge, we can expect that $\tau\rightarrow 1$, thus supporting our estimate of the number of effective configurations.  As can be seen in Tab.~\ref{Tab:taucorr}, the reweighted ensembles typically show a decreased autocorrelation time.  This should be expected because reweighting effectively thins the initial ensemble.  Note this decrease in $\tau_{\rm corr}$ partially compensates for the decrease in $N_{\rm eff}$ arising from the $\overline{\omega}^2/\overline{\omega^2}$ factor inEq.~\eqref{eq:Neff}.

\begin{table}
\label{Tab:taucorr}
\caption{The integrated autocorrelation time defined in Eq.~\eqref{eq:autocorr} for both the unreweighted and reweighted data. Notice that the time is given in units of 5 trajectories for the $m_l=0.02$ ensemble and 10 trajectories for the $m_l=0.01$ ensemble.  The columns are labeled by the light quark mass used to generate the original ensemble.}
\begin{ruledtabular}
\begin{tabular}{ccc|ccc}
unreweighted & $m_l=0.02$ &  $m_l=0.01$ & reweighted & $m_l=0.02$ & $m_l=0.01$ \\
\hline
$P$ & 1.8 & 1.1 &		$P_w$ & 1.1 & 1.0 \\
$Q$ & 7.5 & 3.8 &      $Q_w$ & 1.8 & 1.1 \\
$Q^2$ & 3.8 & 2.5 &    $Q_w^2$ & 1.4  & 1.1 \\
\end{tabular}
\end{ruledtabular}
\end{table}

%gives $Eff=6.3\%$ and $N_{\rm eff}=63$ for the $m_l=0.01$ ensemble, and $Eff=7.9\%$ and $N_{\rm eff}=48$ for the $m_l=0.02$ ensemble. 

\section{Topological Charge and susceptibility}
\label{sec:topology}

The topological charge typically has the longest autocorrelation time among the usual observables and represents long distance physics. We will examine the reweighting effect on it first.  We will later study short distance quantities such as the average plaquette and residual mass.

As the first step in determining the topological charge we use Wilson flow~\cite{Luscher:2010we, Luscher:2010iy} developed by L\"uscher to smooth or cool the gauge configurations.  As a function of the Wilson flow time, the gauge fields obey the following differential equation and initial conditions:
\begin{equation}
  \frac{d U_t(x,\mu)}{dt} = i g_0^2\sum_a \frac{d}{ds} S_w(e^{-i sX^a}U_t)|_{s=0}T^a U_t(x,\mu), \quad U_t(x,\mu)|_{t=0} = U(x,\mu),
\end{equation}
where $S_w(U) = \frac{1}{g_0^2}\sum_p {\rm Re Tr}\{1-U(p)\}$ is the Wilson gauge action and $\{T^a\}_{1 \le a \le 8}$ the eight hermitian generators of $SU(3)$.  When acting on a particular link matrix $U(y,\nu)$ the matrix $X^a = T^a$ if $y=x$ and $\nu=\mu$ and is zero otherwise.  We follow the Runge-Kutta integration scheme (see Appendix C of Ref~\cite{Luscher:2010iy} ).  This realizes a continuous version of the stout smearing technique developed by C. Morningstar and M. Peardon~\cite{Morningstar:2003gk}. We have used an integration time step $\delta t = 0.04$, which gives an error smaller than $10^{-5}$ for $t \le 16$ as seen by comparing the results to those obtained with $\delta t=0.005$ for dozens of configurations. All results given here use $\delta t=0.04$ and the integration error should be negligible for the topological charge calculation of interest.

\begin{figure}[tbh]
\includegraphics[width=80mm]{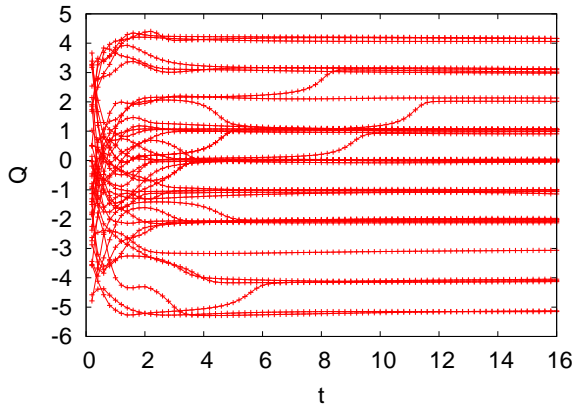}
\caption{Topological charge $Q(t)$ as the function of the Wilison flow time $t$.  We show results from those configurations numbered between 500 and 4500 separated by steps of 100 trajectories from the $m_l=0.01$ ensemble.}
\label{fig:tcharge}
\end{figure}

We then determine the topological charge using the ``5Li'' (5-loop-improved) algorithm~\cite{deForcrand:1997sq} on these ``cooled'' configurations, finding values of topological charge very close to integers.  Figure~\ref{fig:tcharge} shows a few typical evolution curves for the topological charge Q as a function of the flow time t.  Notice that there are ambiguities for some of the configurations: the topological charge converges to one integer, stays there for a while and then moves away converging to a neighboring integer. The problem of defining the topological charge on the lattice and its ambiguities are not the focus of this paper. We will simply adopt the Wilson flow prescription and use the value of $Q$ at $t=16$ as our result for the topological charge.

\begin{figure}[tbh]
  \begin{tabular}{cc}
	\includegraphics[width=80mm]{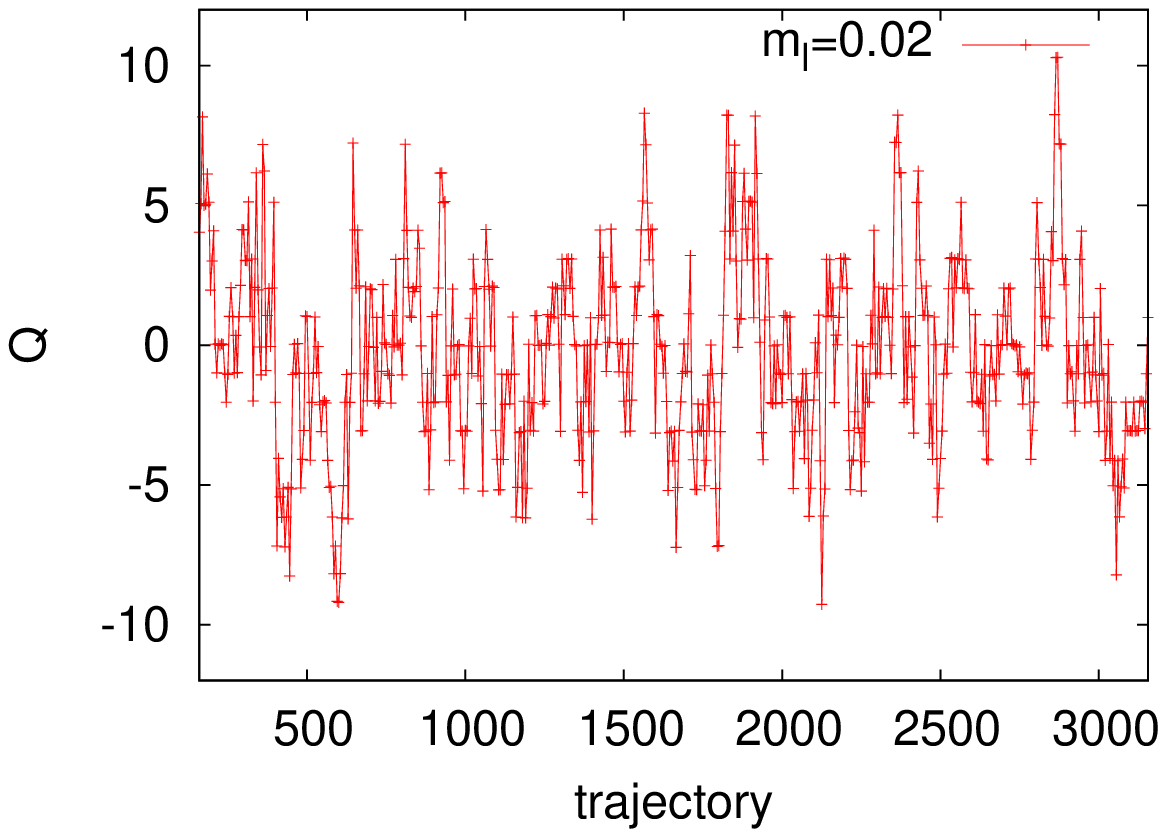} &
	\includegraphics[width=80mm]{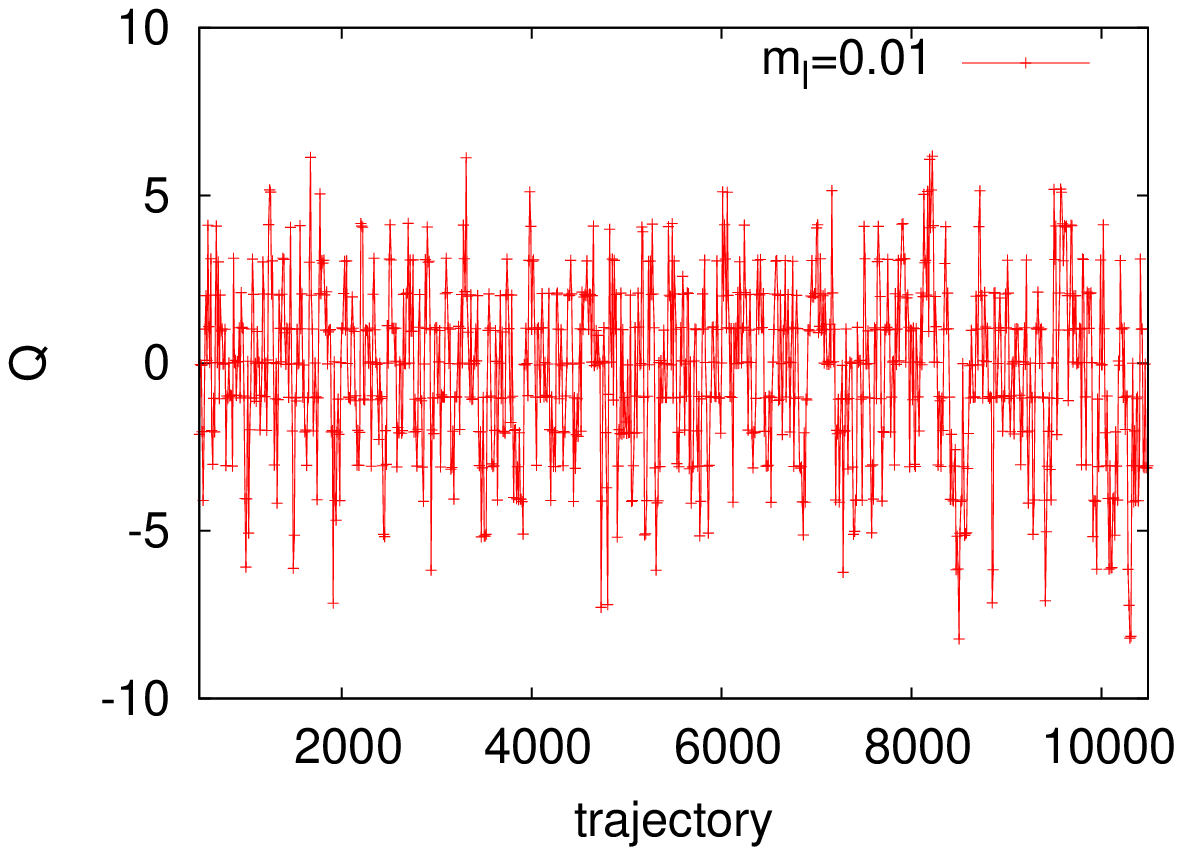} \\
	\includegraphics[width=75mm]{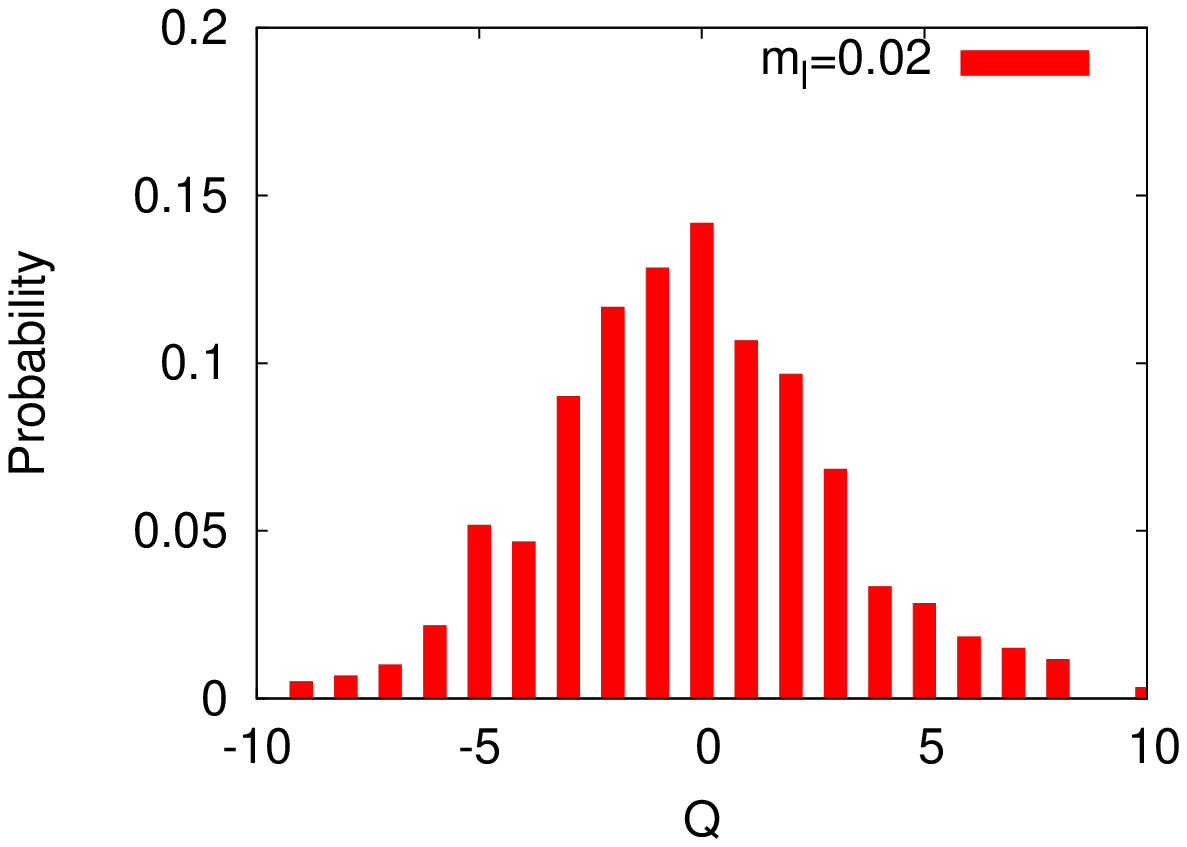} &
	\includegraphics[width=75mm]{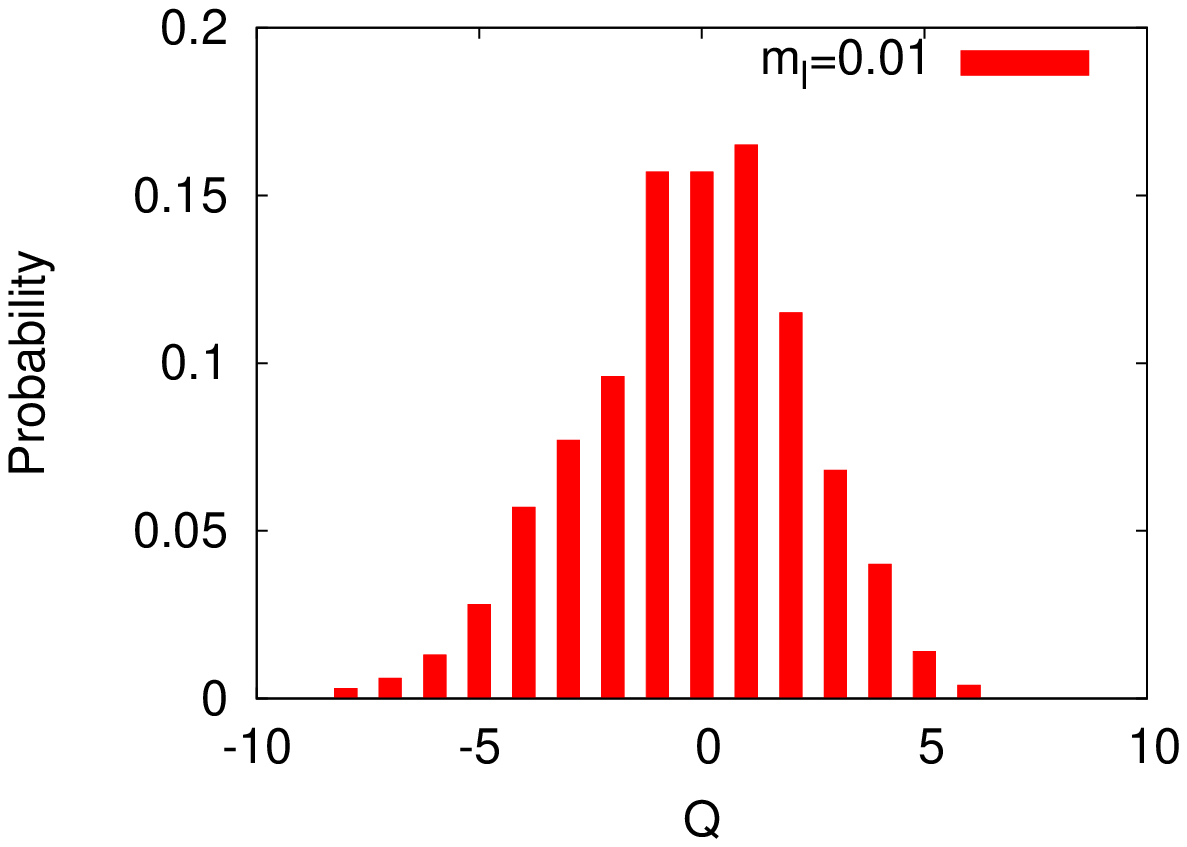} \\
	\includegraphics[width=75mm]{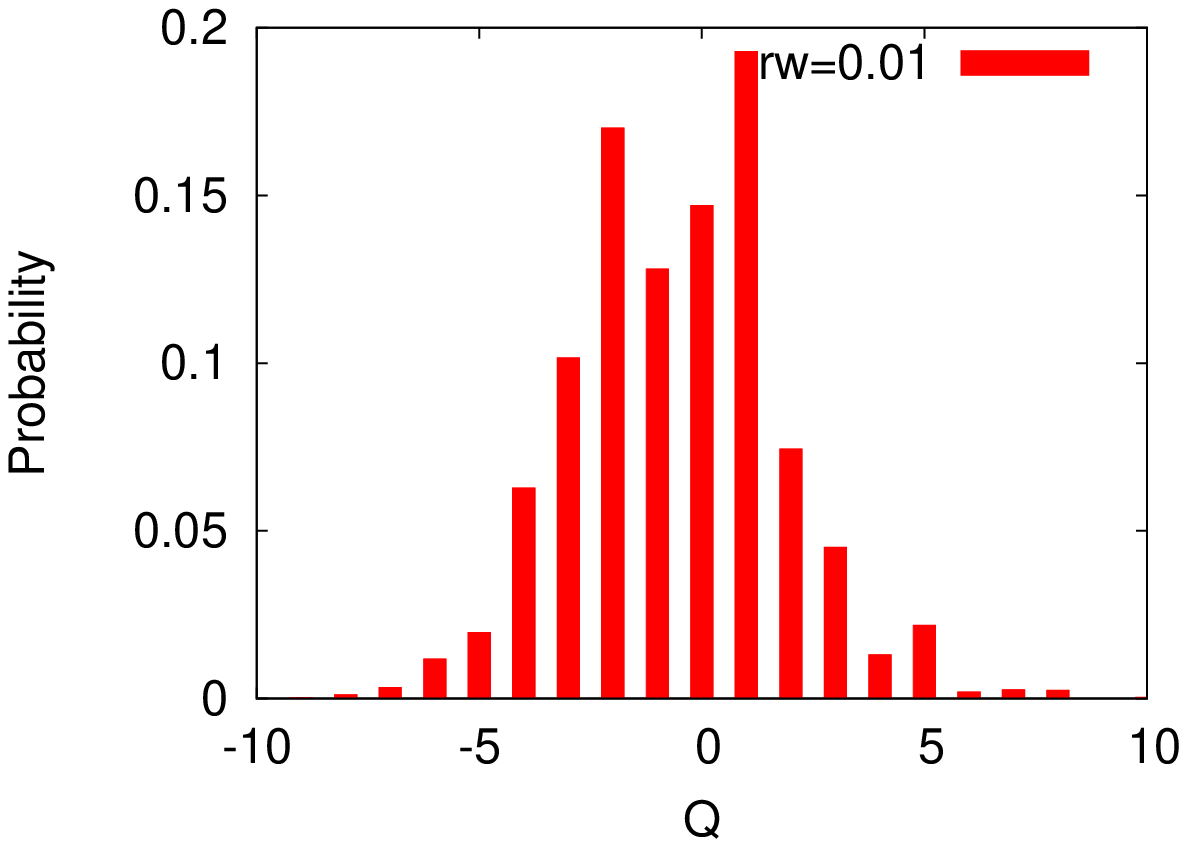} &
	\includegraphics[width=75mm]{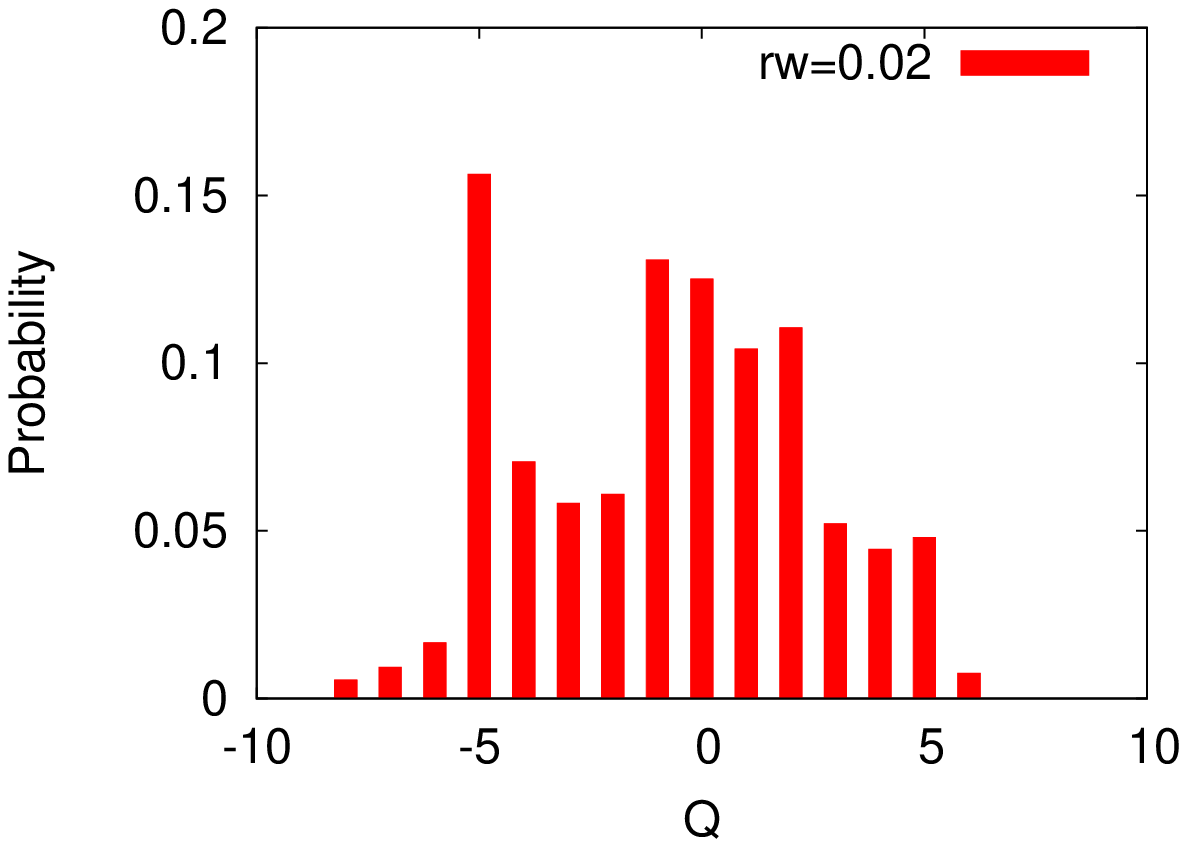} \\
	$\left< Q\right> =-0.25(31)$, $\left<Q^2\right> =10.4(1.3)$ &
	$\left< Q\right> =-0.24(14)$, $\left< Q^2\right> =6.17(38)$ \\
	$\left< Q_w\right> =-0.54(31)$, $\left< Q_w^2\right> =5.9(1.4)$ 
        &$\left< Q_w\right> =-0.76(62)$, $\left< Q_w^2\right> =10.5(2.0)$ 
  \end{tabular}
  \caption{Topological charge evolution (top row) for the $m_l=0.02$ ensemble, measured every fifth trajectory (left), and for the $m_l=0.01$ ensemble, measured at every tenth trajectory (right). 
The middle row shows the resulting topological charge distribution for $m_l=0.02$ (left) and $m_l=0.01$ (right).   The bottom row shows the histograms after reweighting: $m_l=0.02 \to 0.01$ (left) and $m_l=0.01 \to 0.02$ (right) .  Shown below the histograms are the results for $\langle Q\rangle$ and $\langle Q^2\rangle$ for the original and reweighted ensembles.}
\label{fig:Qdistribution}
\end{figure}

Figures~\ref{fig:Qdistribution} (a) and (b) show the evolution of the topological charge for the $m_l=0.02$ and $m_l=0.01$ ensemble respectively. There are apparent autocorrelations among configurations which we have also seen in the corresponding autocorrelation functions shown in Fig.~\ref{fig:autocorr}. The lower two graphs of Fig.~\ref{fig:Qdistribution} present the topological charge distributions and the corresponding distributions obtained after reweighting. The reweighted topological charge distribution is given by the probability function $P_w(Q)$ computed from the relation: 
 \begin{equation}
P_w(Q) = \frac{\sum_{i'=1}^N w_{i'}\delta_{Q_{i'},Q} }{\sum_{i=1}^N w_i} 
\end{equation}
where the Kronecker delta factor $\delta_{Q_{i'},Q}$ ensures that sum over $i'$ only includes those configurations having topological charge Q. 

The most important property of the topological charge distribution (given its central value of zero) is the width $\left<Q^2\right>$ of the distribution, which is related to the topological susceptibility. The calculated values of $\left<Q^2\right>$ with the average $\left<Q\right>$ and the reweighted values  $\left<Q_w^2\right>$ and $\left<Q_w\right>$ are summarized below the  distributions shown in Fig.~\ref{fig:Qdistribution}. The quoted errors are calculated from the fluctuations seen among blocks of 100 trajectories to remove any effect of autocorrelation. 

We can see very clearly how we have successfully reweighted one distribution to another. If we reweight the light sea quark mass from $0.02$ to $0.01$, the resulting $m_l-0.01$ ensemble reweighted from $m_l=0.02$ gives a distribution with second moment  $\left<Q_w^2\right>=5.9(1.4)$, in excellent agreement with the `true' value from our dynamically generated $m_l=0.01$ ensemble, which gives $\left<Q^2\right>=6.17(38)$.  In fact, a detailed comparison of the two $m_l=0.01$ ensembles in Fig.~\ref{fig:Qdistribution} (the reweighted histogram in the bottom row, left column and the directly computed histogram in the middle row, right column) shows very similar features.  Comparing the tails of the distribution  one clearly sees that the configurations with large topological charge are strongly suppressed by the reweighting factors: the reweighting really reflects the underlying physics by giving small weights to large topological charge configurations and large weights to small topological charge configurations. Similarly, the reweighting from the $m_l=0.01$ ensemble to that with $m_l=0.02$ is also successful but looks less impressive as can be seen by comparing the reweighted histogram in the bottom row, right column and the directly computed histogram in the middle row, left column. This should be expected as we try to reweight a narrower distribution to a wider one.  The reweighting must give large weights to the configurations with large topological charge, which are rare in the $m_l=0.01$ ensemble.  Thus, in the reweighted distribution the far ends of the tails which are populated in the directly computed ensemble are missing from the reweighted one: these bins are empty in our finite sample and it is not possible to create something by reweighting. This may suggest that it is more favorable to reweight from heavier mass to lighter mass. 

The spike at $Q=-5$ in the $m_l=0.02$ distribution reweighted from the $m_l=0.01$ ensemble in Fig.~\ref{fig:Qdistribution} (bottom row, right-hand column) may look troublesome. We find that its contribution comes to a large extent from a single configuration that has the largest reweighting factor in Fig.~\ref{fig:rwfactor}. Therefore, this large spike has very large inherent uncertainly. To evaluate the quality of the data set, we (incorrectly) dropped this configuration and performed a complete analysis, repeating all the calculations in this paper.   This experiment revealed no surprises, giving results consistent with those presented here which included all configurations. Nonetheless, the unphysical topological charge distribution for the ensemble reweighted to $m_l=0.02$ ensemble suggests that our calculation is not far from being sensitive to uncontrolled statistical fluctuations. The reweighting is not ideal.  It would be prudent either to have more configurations or to reweight less aggressively in the mass difference. 

\section{Average plaquette}
\label{sec:plaquette}

\begin{figure}
  \includegraphics[width=80mm]{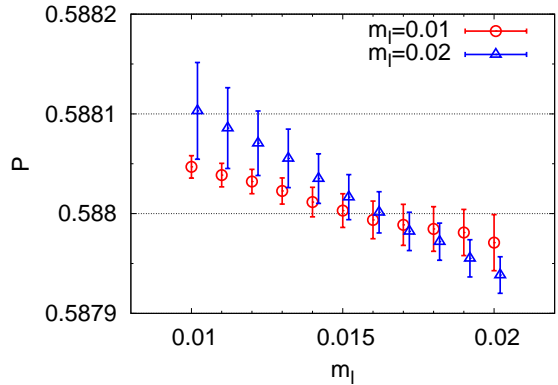} 
  \caption{Reweighted values for the average plaquette as the dynamical light quark mass is decreased from $m_l=0.02$ to $m_l=0.01$ using the $m_l=0.02$ ensemble (open circles).  Also shown are results from the $m_l=0.01$ ensemble as $m_l$ is increased by reweighting from $m_l=0.01$ to $m_l=0.02$. The data from the $m_l=0.02$ ensemble are slightly offset in order to show the overlapped data points more clearly.  While each reweighting sequence was performed in 80 steps, we plot only the result of each eighth step of 0.001.}
  \label{fig:plaq}
\end{figure}

Now let us discuss how well reweighting works when applied to a representative short distance quantity, the average plaquette. The reweighting is performed in 80 steps, so we can examine all the intermediate reweighted results. We plot the reweighted results from each mass change of 0.001 for reweighting in both directions in Fig.~\ref{fig:plaq}.  As we can see, the average plaquette is successfully reweighted from 0.587938(18) for the $m_l=0.02$ ensemble to the 0.588103(49) value predicted for $m_l=0.01$ as well as from the 0.588047(11) for the $m_l=0.01$ ensemble back to 0.587971(28) value predicted for $m_l=0.02$.  The results, including all the intermediate mass values, agree within $1 \sigma$. The quoted error is calculated using a block size of 50 trajectories to remove any effect of autocorrelation.  We will also use a block size of 50 trajectories to calculate the error in our later calculation of $m_{\rm res}$, $f_\pi$ and $m_\pi$.  At this point, we have shown that light-quark mass reweighting is very successful for both long distance and short distance physics. 

\begin{figure}
  \begin{tabular}{cc}
  	\includegraphics[width=80mm]{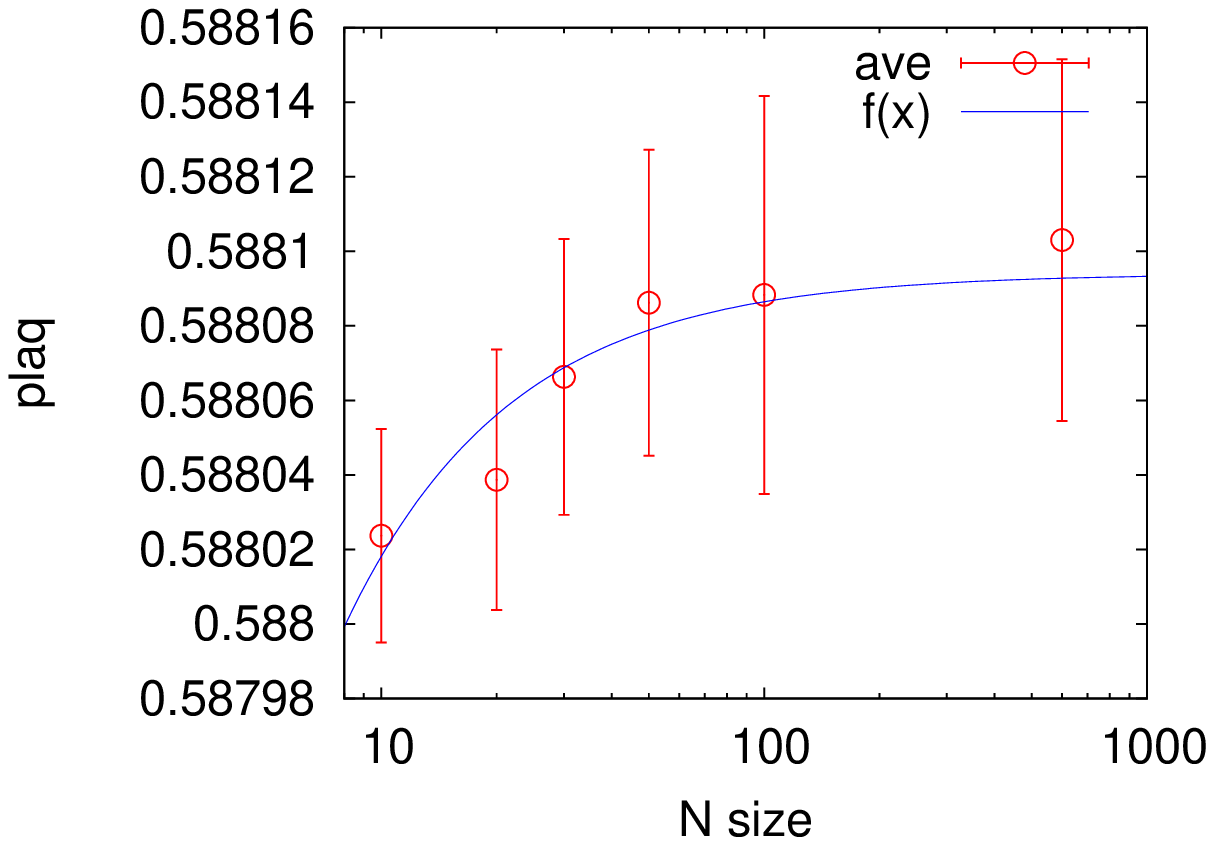} &
	\includegraphics[width=80mm]{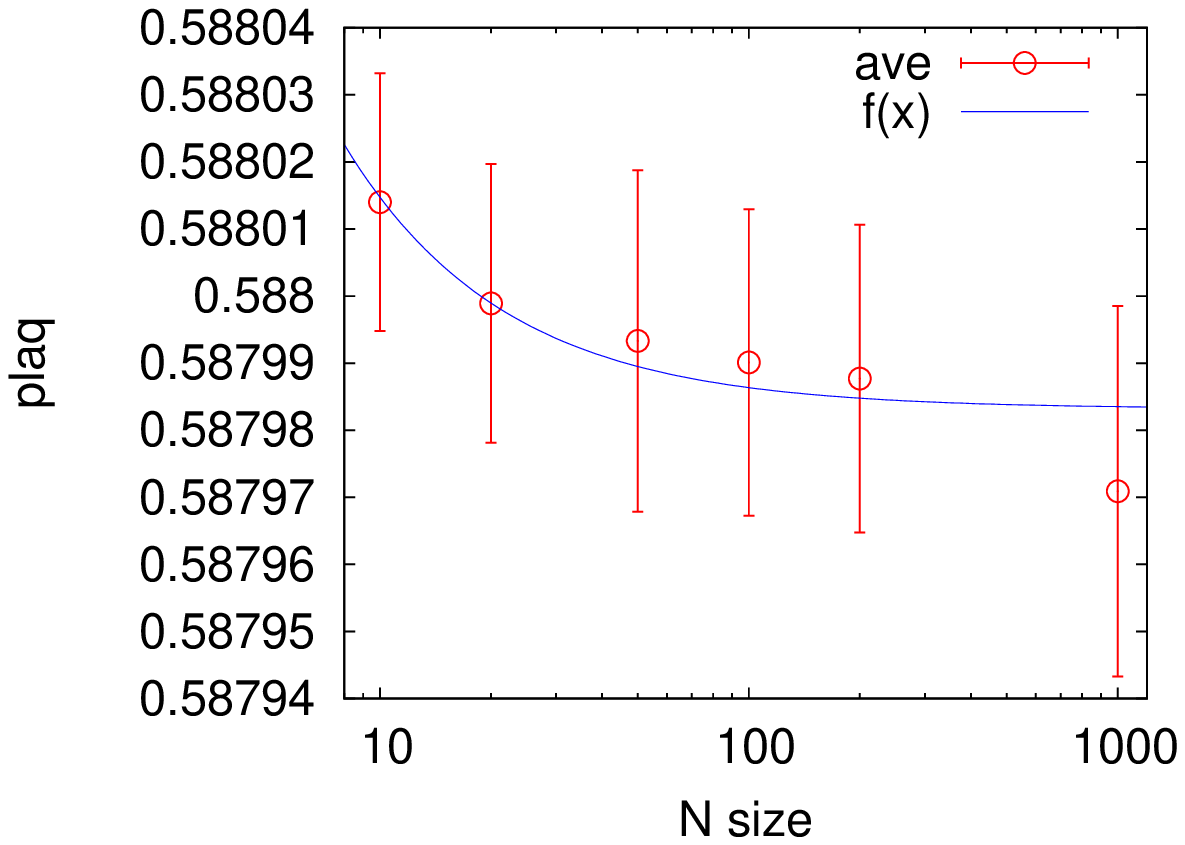} \\
 \end{tabular}
 \caption{Biased estimator of the reweighted average plaquette. For $m_l=0.02$ ensemble(left), we choose sample size to be N=10, 20, 30, 50, and 100. The corresponding number of measurements are 60, 30, 20, 12 and 6. For $m_l=0.01$ ensemble(right), we choose sample size to be N=10, 20, 50, 100, and 200. The corresponding number of measurements are 100,50,20,10 and 5. The data is then fit with function $f(x)=P_w(1+c/N)$.}
 \label{fig:bias}
\end{figure}

So far we have ignored the bias which arises from the finite sample size used in our stochastic estimate of the reweighted quantity $\langle O_w\rangle_N$ given by the ratio in Eq.~\eqref{eq:OwN}.  Let us take the average plaquette as an example and see how large it is. Suppose the sample size is N.  The reweighted result $\left<P_w\right>_N$ is biased in the sense that the expectation value $\left<\!\left<\left<P_w\right>_N\right>\!\right>$ will not agree with the true value $\overline{P}_w= \lim_{N\to\infty}\left<P_w\right>_N$. It is straightforward to work out the difference, keeping only the lowest order terms:
\begin{eqnarray}
\left<\!\left<\left<P_w\right>_N\right>\!\right> & = & \left<\!\left<\frac{\sum_{i=1}^N w_i P_i}{\sum_{j=1}^N w_j}\right>\!\right> \\
& \approx & \overline{P}_w\Biggl\{1+\frac{1}{N^2}\frac{1}{\overline{w}^2\overline{P}_w} \sum_{i=1}^N  \sum_{l=1}^N \Bigl\langle\!\Bigl\langle w_i w_{l}(\overline{P}_w - P_i)\Bigr\rangle\!\Bigr\rangle \Biggr\}.
\end{eqnarray}
We can see that the bias is proportional to $1/N$, and the coefficient is related to the correlation of $w_i$ and $P_i$.   We might attempt to estimate the bias given in this equation for our single sample by omitting the average over samples, $\langle\!\langle\ldots\rangle\!\rangle$ and by examining the $l=i$ term which might be expected to be dominant.   This term can be written as $-\sum_{i,j=1}^Nw_iw_j(w_i-w_j)(P_i-P_j)/2N^3\overline{w}^3$.  This expression hints that the bias is negative for a positive correlation between $w_i$ and $P_i$, and vice versa.  As should be expected, if there is no correlation between $P_i$ and $w_i$, the coefficient vanishes. Taking all of our data and dividing them into small samples, we calculate the expectation value of $\left<P_w\right>_N$. The results are shown in Fig.~\ref{fig:bias} and suggest that the bias for the $m_l=0.02$ ensemble is negative and the bias for the $m_l=0.01$ ensemble is positive although these effects are at or below the one $\sigma$ level. These results also suggest that the bias is negligible for reasonably large ensemble size $N>20$. 

\section{Pion Mass and Decay Constant}
\label{sec:pion_fpi_mres}

Let us next study the effect of reweighting on some additional, interesting physical quantities. In this section, we will examine the sea quark dependence of the residual mass, the pion mass, and the pion decay constant. In the following, for the $m_l=0.02 (0.01)$ ensemble, we fix the valence quark mass at 0.01(0.02) and gradually reweight the sea quark mass from 0.02(0.01) to 0.01(0.02), and study the change in each of these physical quantities. Once each measurements has been reweighted to $m_l=0.01(0.02)$, it becomes unitary with equal valence and sea light quark masses and we compare the reweighted result with the unitary result directly calculated from the $m_l=0.01(0.02)$ ensemble.  Note we will express all quantities in lattice units unless other units are explicitly specified.

\begin{figure}[!tbh]
  \includegraphics[width=80mm]{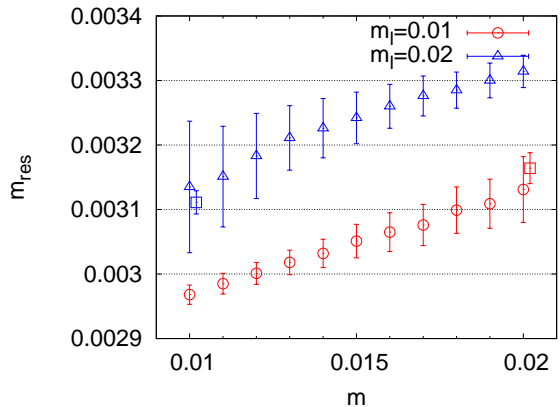} 
  \caption{Reweighted results for the residual mass. The blue triangular (red circle) points are the residual mass calculated with $m_l=0.02(0.01)$ ensemble, using valence quark mass 0.01(0.02). The two box-shaped points give the results from a direct, unitary calculation with $m_{\rm val}=0.02$ on $m_l=0.02$ ensemble, and $m_{\rm val}=0.01$ on $m_l=0.01$ ensemble. They are intentionally shifted by a small amount horizontally to avoid overlap with the reweighted results.}
  \label{fig:mres}
\end{figure}

The residual mass is an important quantity in the DWF formulation, representing a contribution to the physical quark mass from the explicit chiral symmetry breaking which is present if the lattice extent in the fifth dimension is finite.  (For this calculation the fifth dimensional extent has the value $L_s=16$.)  The residual mass is an interesting quantity to study because it typically shows a strong dependence on the light sea quark mass and so should provide an interesting test of light sea quark mass reweighting.   Our results obtained using the methods of Ref.~\cite{Allton:2008pn} for the residual mass from both the $m_l=0.02$ and $m_l=0.01$ ensembles are shown in Fig.~\ref{fig:mres}.  This figure shows that the residual mass is positively correlated with the light sea quark mass (at fixed valence quark mass).   We see that the ensemble with sea quark mass $m_l=0.02$  when reweighted to $m_l=0.01$ gives a result that is consistent with that obtained by direct calculation on the $m_l=0.01$ ensemble, and vice versa.
 
To obtain $m_\pi$ and $f_\pi$, we calculated three different correlation functions. Using a gauge-fixed pseudo-scalar wall source, we choose the sink to be either a gauge-fixed pseudo-scalar wall sink, a pseudo-scalar point sink or a point sink formed from the zeroth component of the axial current. These correlation functions are identified as $C_{O_1O_2}^{{\rm sink}_1{\rm source}_2}(t)$ which for our three cases becomes $C_{PP}^{WW}(t)$, $C_{PP}^{LW}(t)$ and $C_{AP}^{LW}(t)$. Here the label `W' means a gauge-fixed wall source or sink, `L' means a point sink, `P' means pseudo-scalar operator, 'A' means the zeroth component of the axial current operator. We perform a simultaneous fit to all three correlators at large times, using the functional form
\begin{equation}
C_{O_1O_2}^{s_1s_2}(t) = N_{O_1O_2}^{s_1s_2} [e^{-m_\pi t} + e^{-m_\pi(T-t)}]
\end{equation}
This fit give us $m_\pi$ directly. From the amplitudes $N_{O_1O_2}^{s_1s_2}$, we construct $f_\pi$ following Ref~\cite{Aoki:2010dy}, 
\begin{equation}
f_\pi = Z_A\sqrt{\frac{2}{m_\pi}\frac{(N_{AP}^{LW})^2}{N_{PP}^{WW}}}
\end{equation}
where $Z_A$ is the renormalization constant of the local axial current $A_\mu$, calculated in Ref.~\cite{Allton:2007hx}.  Since $Z_A$ is to be evaluated in the chiral limit and thus does not depend on sea quark mass, we simply set $Z_A=1$ for the present calculation.
 
 \begin{figure}[!tbh]
  \begin{tabular}{ll}
  \includegraphics[width=80mm]{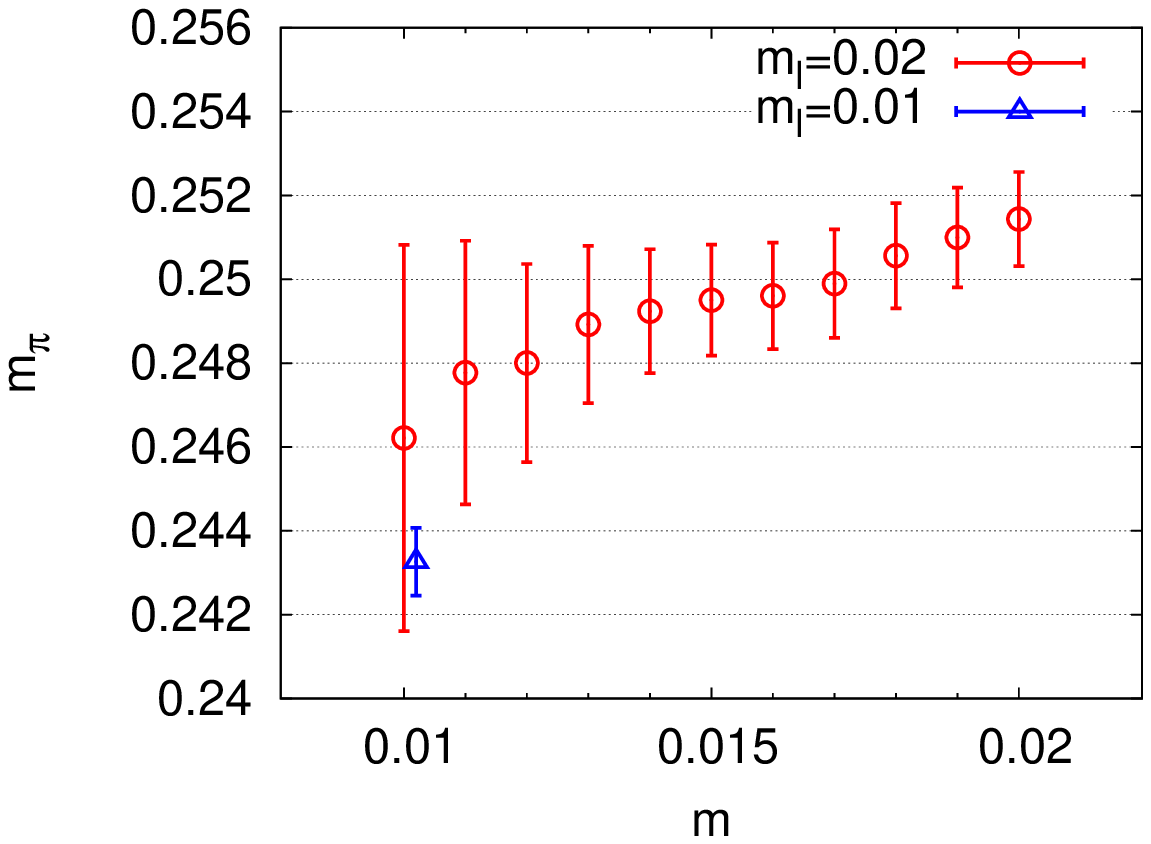} &
  \includegraphics[width=80mm]{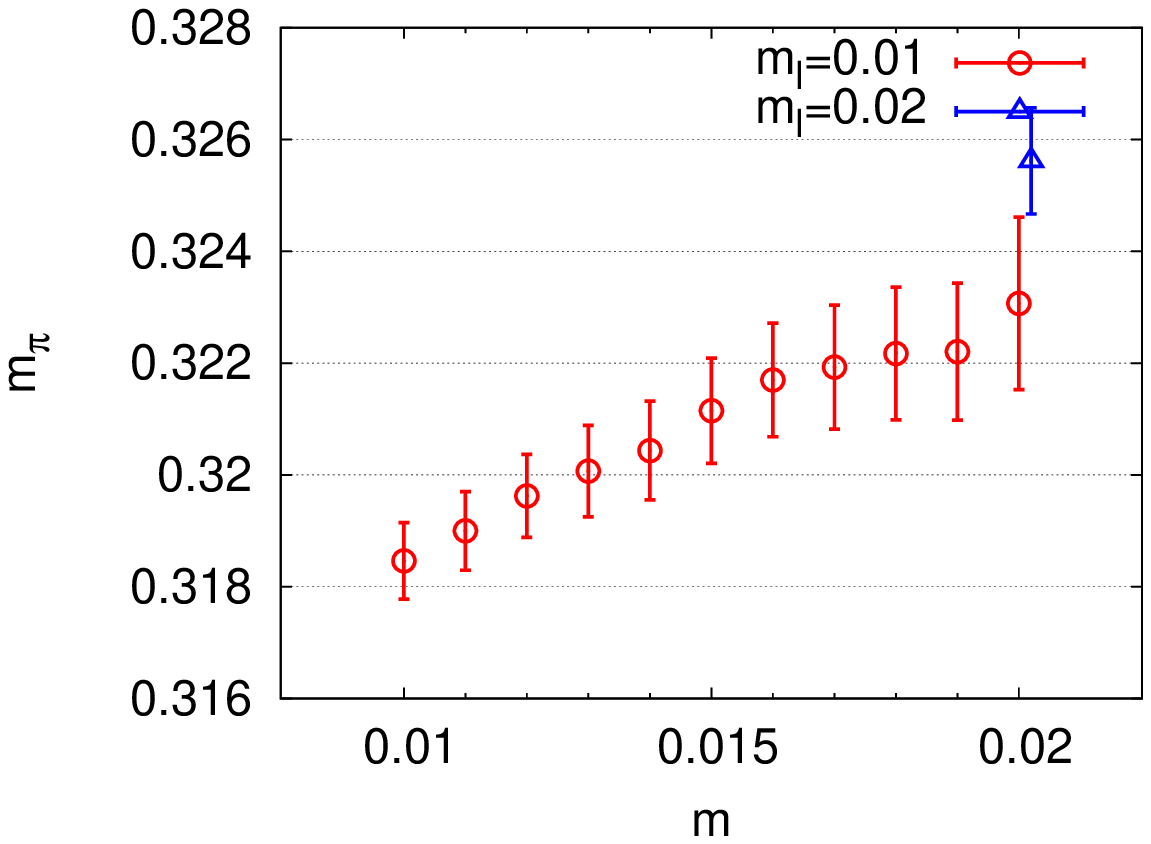} \\
  \end{tabular}
  \caption{The pion mass computed on the $m_l=0.02$ ensemble (left panel) at a fixed valence quark mass $m_{\rm val} = 0.01$ as a function of the reweighted light sea quark mass with the results shown as open circles.  The right panel shows similar results for the pion mass computed using the
$m_l=0.01$ ensemble for fixed $m_{\rm val} = 0.02$.   The triangular points in each panel show the result computed on the other ensemble with equal sea and valence quark masses: $m_l=0.01$ (left panel) and $m_l=0.02$ (right panel).}
  \label{fig:pionmass}
\end{figure}

\begin{figure}[tbh]
  \begin{tabular}{ll}
  \includegraphics[width=80mm]{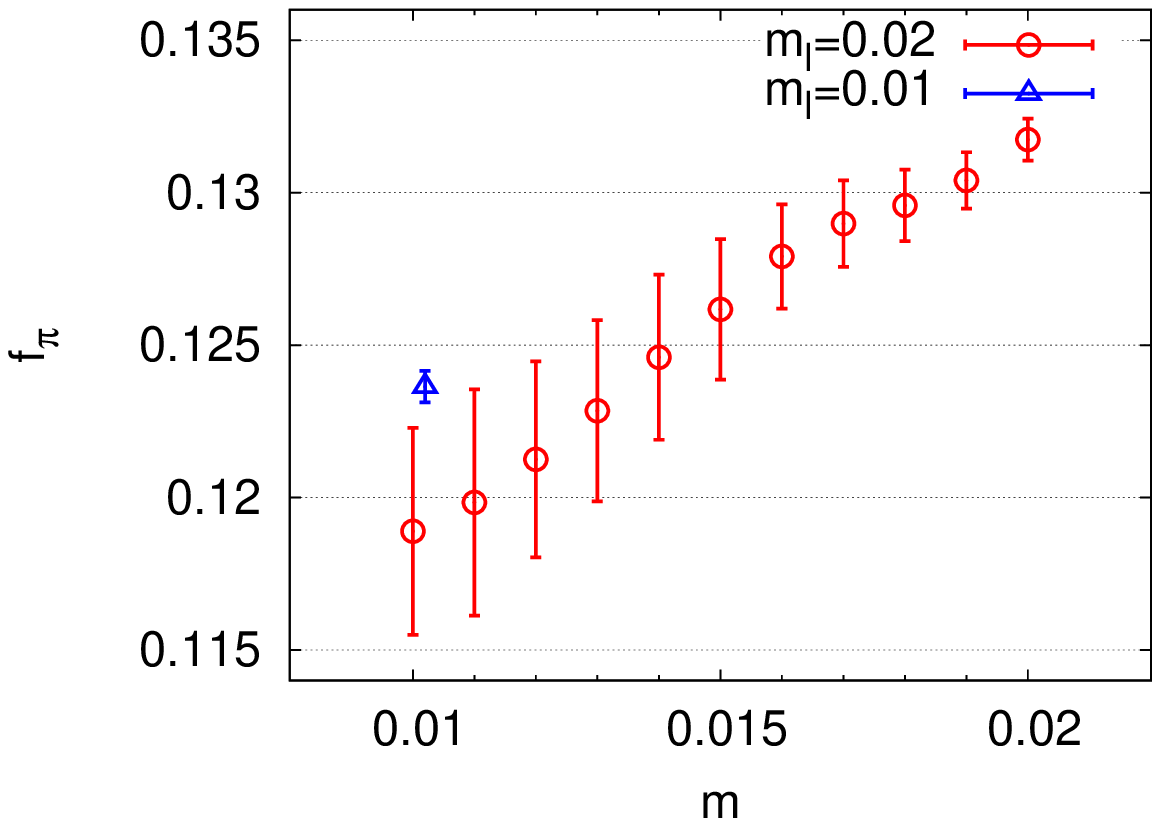} &
  \includegraphics[width=80mm]{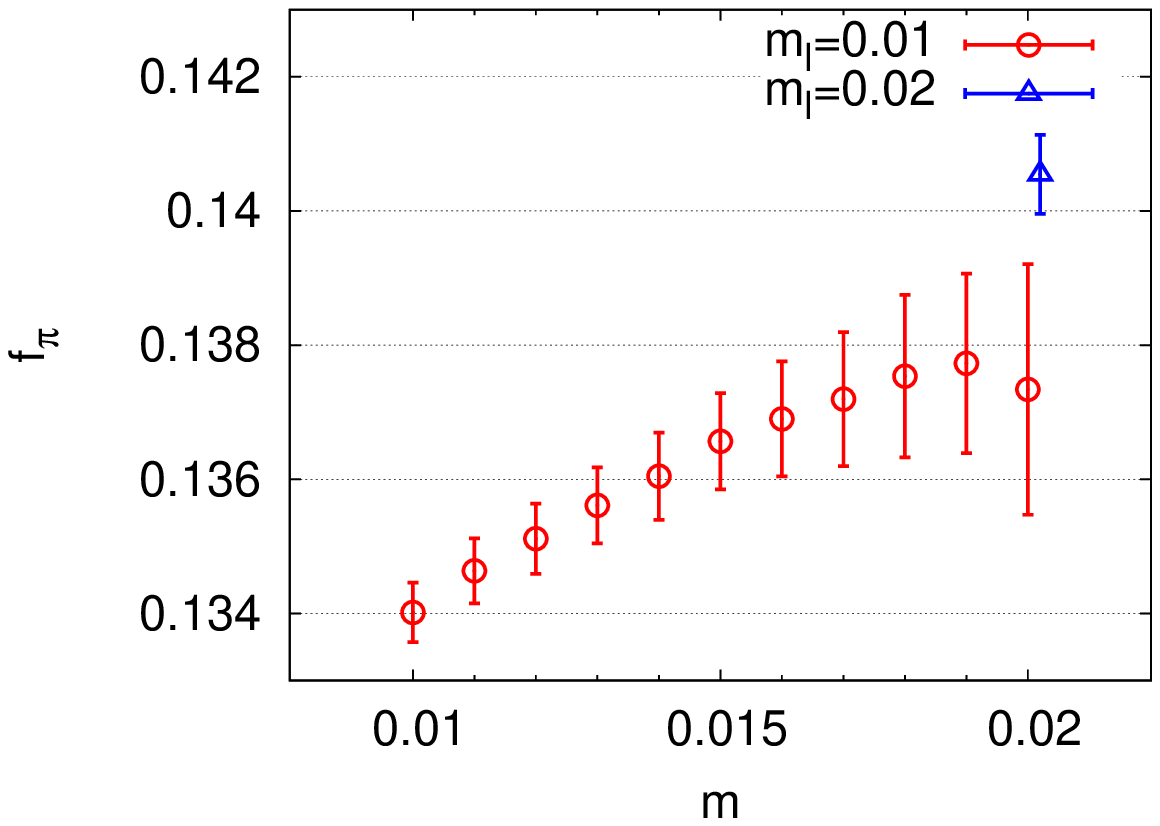} \\
  \end{tabular}
  \caption{The reweighted pion decay constant plotted in the same style as used in Fig.~\ref{fig:pionmass}.}
  \label{fig:fpi}
\end{figure}
 
 Figure~\ref{fig:pionmass} shows the calculated dependence  of the pion mass on the light sea quark mass. In the left graph, we show the reweighting of the $m_l=0.02$ ensemble, using a fixed valence quark mass $m_{val}=0.01$. The results from the intermediate reweighting steps are also shown. Similarly, the right graph shows the reweighting on the $m_l=0.01$ ensemble. These results demonstrate a positive dependence of the pion mass on the light sea quark mass. The reweighted results show a clear trend, although with increasing errors, as we reweight the sea quark mass.  In both cases the result obtained in the final step of the reweighting procedure agrees with the directly calculated result at the 1 $\sigma$ level. Figure.~\ref{fig:fpi} shows similar results for $f_\pi$, presented in the same way as the data for $m_\pi$. in Fig.~\ref{fig:pionmass}.  Similar conclusions can be drawn.
 
 \section{Scalar correlator}
\label{sec:correlator}

Finally, let us look at something less successful but equally interesting. The properties of the iso-vector, scalar particle are known to be difficult to calculate because its mass is much larger than the mass of the pion, so large are statistics are needed. For the reweighted data, the effective number of configurations drops to less than one hundred, resulting in such a small ensemble size that it is not possible to determine the mass of this scalar state.  However, by simply looking at the scalar correlator itself, we can see the dramatic effect of reweighting: the partially quenched violation of positivity disappears as the sea quark mass is reweighted to the unitary value.

\begin{figure}[!tbh]
  \begin{tabular}{cc}
  \includegraphics[width=80mm]{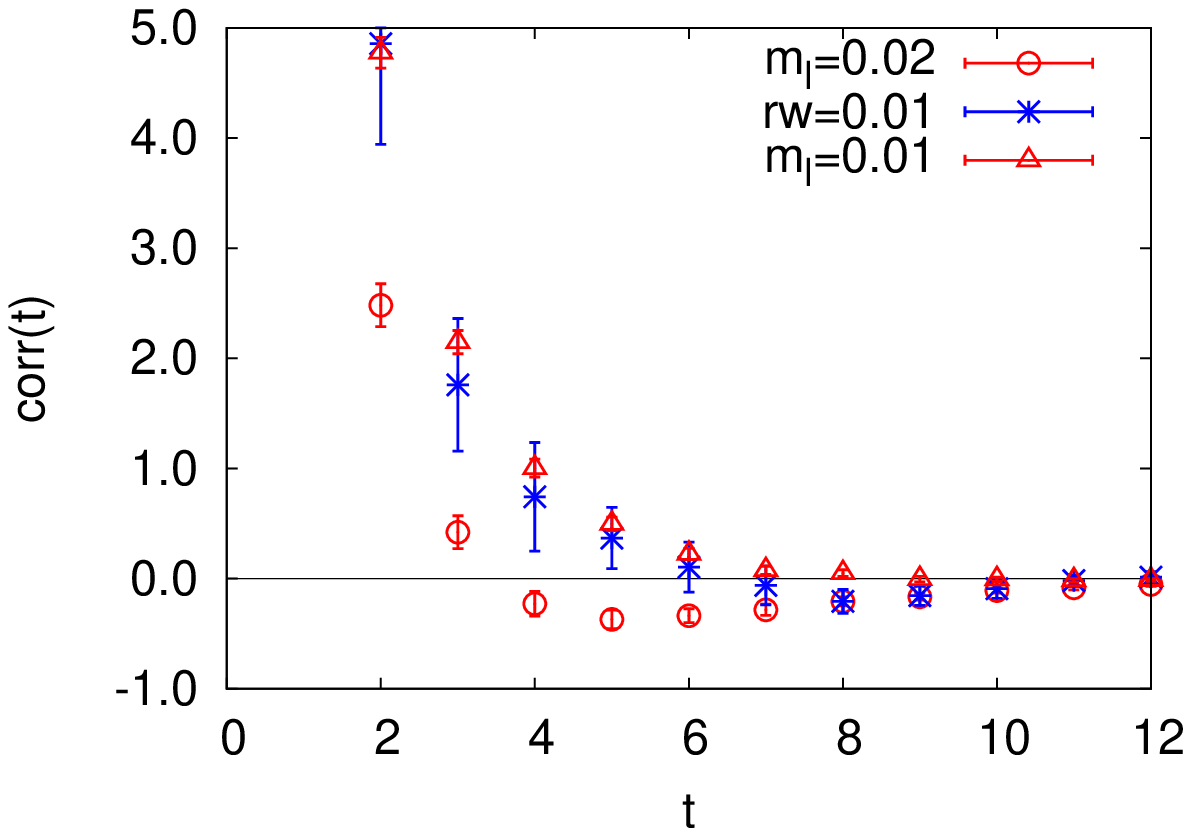} &
  \includegraphics[width=80mm]{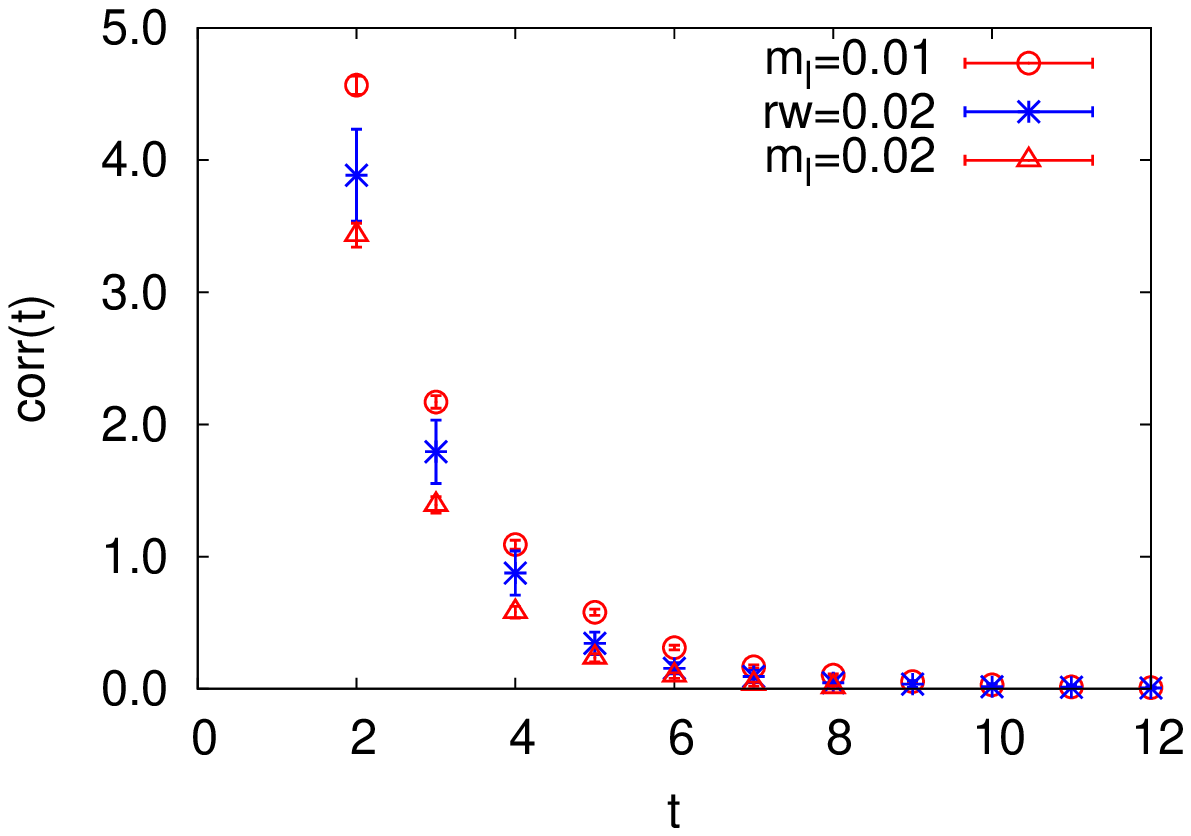} \\
  $m_{\rm val}=0.01$  & $m_{\rm val}=0.02$
  \end{tabular}
  \caption{Reweighting results for the scalar correlation function plotted as a function of time. The left panel shows this propagator computed with a valence quark mass of 0.01.  The open circles give the non-unitary result from 0.02 ensemble, the crosses the same observable reweighted to the unitary, light sea quark mass of 0.01 and the triangles the result of a direct calculation on the $m_l=0.01$ ensemble.  The right panel shows the corresponding reweighting on the $m_l=0.01$ ensemble.}
  \label{fig:scalar}
\end{figure}

We  calculate the scalar correlator using an identical, gauge-fixed wall source and sink so that
the resulting correlation function should be positive. A well known artifact of partial quenching appears in the case where $m_{val}<m_{sea}$: the correlator becomes negative for large source sink time separations when the lighter, negative norm $\pi\eta^\prime$ intermediate state dominates~\cite{Bardeen:2001jm}.  This behavior can be clearly seen in the left panel of Fig.~\ref{fig:scalar} where the valence quark mass is 0.01, and the sea quark mass is 0.02. The correlator becomes negative for $t>4$.   After reweighting the sea quark mass from 0.02 down to the unitary value of 0.01, the correlator data points increase and becomes positive as seen previously in Ref.~\cite{Hasenfratz:2008fg}.  In the left-hand panel one can also see the result from a unitary calculation performed directly with $m_{\rm val}=0.01$ on the $m_l=0.01$ ensemble.  These results agree within the larger statistical errors of the reweighted data.  In contrast, the partially quenched result with $m_{\rm val}=0.02$ determined on the $m_l=0.01$ ensemble gives a positive value for the correlator as shown in the right panel of Fig.~\ref{fig:scalar}. The reweighting from $m_{sea}=0.01$ to 0.02 now has a less dramatic effect but does give a result in agreement with a direct calculation with  $m_{\rm val}=0.02 $ on the $m_l=0.02$ ensemble.  Because of the poor statistics, the comparison between the reweighted and direct results is not accurate, especially at large $t$ where the relative error increases dramatically and the results become unreliable.

\section{Summary and discussion}
\label{sec:summary}

We have shown how reweighting can be used to vary the light sea quark mass while working with a single ensemble generated with a single sea quark mass. We first successfully reweighted the topological charge distribution from one of our light sea quark masses to the another. We found that it is better to reweight from a heavier mass to a lighter mass, because this corresponds to reweighting from a wider distribution to one that is more narrow.  In this case, the poor statistics present in the tail of the distribution is suppressed instead of magnified by the reweighting factor.  Next we studied the reweighting of the average plaquette, examining results obtained for a sequence of intermediate masses.  In addition, we examined the bias introduced by our stochastic finite-sample  reweighted measurement of the average plaquette. We found that the bias is negligible for fairly large ensemble with $N>20$. After this study of both long distance and short distance quantities, we examined the behavior under reweighting of the residual mass, pion mass and pion decay constant.  In each case, reweighting revealed how the given physical quantity varied with the sea quark mass when computed on an ensemble with a fixed quark ass.  In each case, the final step of reweighting was chosen to duplicate the sea quark mass for which had been used to generate a second ensemble.  This allowed a direct comparison between the reweighted and a directly generated result and in each case the two results agreed within errors.  Finally, reweighting was shown to dramatically to correct the partial quenching artifacts seen in the calculation of the iso-scalar, scalar correlator initially computed with $m_{\rm val} = 0.01 $ on an $m_l=0.02$ ensemble.

The error on the quantities we have studied was generally increased 3-4 times by reweighting, in rough agreement with our estimate of the number $N_{\rm eff}$ of effective configurations.  These estimates suggested an error increase of a factor of $\sqrt{600/48}=3.5$ and $\sqrt{1000/62} =4$, if autocorrelation is neglected. The topological charge shows the longest autocorrelation times and the associated errors do not follow our estimated value of $N_{\rm eff}$.  However, the relative error on $Q^2$ increases by a factor of 1.5 when reweighting the $m_l=0.02$ ensemble and by 3 for the $m_l=0.01$ ensemble.  As explained in the discussion Sec.~\ref{sec:method}, this possibly smaller increase in error may result from there being fewer independent configurations in the original ensemble so that less information is lost when a portion of these correlated samples is assigned a small weight.

It should be emphasized that the change in the quark mass from $m_l=0.02$ to 0.01 and the reverse is a large change to be accomplished by reweighting.  This is certainly suggested by the large fluctuations of more than 4 orders of magnitude seen in the reweight factors. The resulting efficiency in the reweighting procedure, $N_{\rm eff}/N$, is less than 10\%, causing the error after reweighting to increase by a factor of 4.   However, such an aggressive use of reweighting may be appropriate for a test of the method, more clearly revealing its power and potential limitations.

In real applications of reweighting, we may choose to be more conservative and reweight over a smaller range of mass than considered here to keep the efficiency high.  Had we only reweighted by half of the mass difference, reweighting both ensembles to $m_l=0.015$, the effective number of configurations would increase dramatically from 48 to 308 for the $m_l=002$ ensemble, and from 63 to 386 for the $m_l=0.01$ ensemble. Depending on the physics being studied, we must ensure that $N_{\rm eff}$ is large enough to give statistically useful results. For example,  $N_{\rm eff}<100$, is apparently insufficient for a study of the iso-vector, scalar particle mass.  In conclusion, reweighting has been shown to work remarkably well, even for a large change in the light quark mass, with the resulting errors well described by an effective number of configurations given in Eq.~\ref{eq:Neff}.

\begin{acknowledgments}
We thank our RBC/UKQCD collaborators for encouragement, helpful discussions and suggestions. We acknowledge the RIKEN BNL Research Center and the Brookhaven National Laboratory for providing the facilities on which this work was performed. Qi Liu acknowledges the support of a DOE High Energy Theory Fellowship. Norman Christ and Qi Liu were supported in part by U.S.~DOE grant DE-FG02-92ER40699 and Chulwoo Jung by U.S.~DOE grant AC-02-98CH10886.
\end{acknowledgments}

\bibliography{citations}

\begin{thebibliography}{16}%
\makeatletter
\providecommand \@ifxundefined [1]{%
 \@ifx{#1\undefined}
}%
\providecommand \@ifnum [1]{%
 \ifnum #1\expandafter \@firstoftwo
 \else \expandafter \@secondoftwo
 \fi
}%
\providecommand \@ifx [1]{%
 \ifx #1\expandafter \@firstoftwo
 \else \expandafter \@secondoftwo
 \fi
}%
\providecommand \natexlab [1]{#1}%
\providecommand \enquote  [1]{``#1''}%
\providecommand \bibnamefont  [1]{#1}%
\providecommand \bibfnamefont [1]{#1}%
\providecommand \citenamefont [1]{#1}%
\providecommand \href@noop [0]{\@secondoftwo}%
\providecommand \href [0]{\begingroup \@sanitize@url \@href}%
\providecommand \@href[1]{\@@startlink{#1}\@@href}%
\providecommand \@@href[1]{\endgroup#1\@@endlink}%
\providecommand \@sanitize@url [0]{\catcode `\\12\catcode `\$12\catcode
  `\&12\catcode `\#12\catcode `\^12\catcode `\_12\catcode `\%12\relax}%
\providecommand \@@startlink[1]{}%
\providecommand \@@endlink[0]{}%
\providecommand \url  [0]{\begingroup\@sanitize@url \@url }%
\providecommand \@url [1]{\endgroup\@href {#1}{\urlprefix }}%
\providecommand \urlprefix  [0]{URL }%
\providecommand \Eprint [0]{\href }%
\providecommand \doibase [0]{http://dx.doi.org/}%
\providecommand \selectlanguage [0]{\@gobble}%
\providecommand \bibinfo  [0]{\@secondoftwo}%
\providecommand \bibfield  [0]{\@secondoftwo}%
\providecommand \translation [1]{[#1]}%
\providecommand \BibitemOpen [0]{}%
\providecommand \bibitemStop [0]{}%
\providecommand \bibitemNoStop [0]{.\EOS\space}%
\providecommand \EOS [0]{\spacefactor3000\relax}%
\providecommand \BibitemShut  [1]{\csname bibitem#1\endcsname}%
\let\auto@bib@innerbib\@empty
%</preamble>
\bibitem [{\citenamefont {Aoki}\ \emph
  {et~al.}(2011{\natexlab{a}})\citenamefont {Aoki}, \citenamefont {Arthur},
  \citenamefont {Blum}, \citenamefont {Boyle}, \citenamefont {Brommel} \emph
  {et~al.}}]{Aoki:2010pe}%
  \BibitemOpen
  \bibfield  {author} {\bibinfo {author} {\bibfnamefont {Y.}~\bibnamefont
  {Aoki}}, \bibinfo {author} {\bibfnamefont {R.}~\bibnamefont {Arthur}},
  \bibinfo {author} {\bibfnamefont {T.}~\bibnamefont {Blum}}, \bibinfo {author}
  {\bibfnamefont {P.}~\bibnamefont {Boyle}}, \bibinfo {author} {\bibfnamefont
  {D.}~\bibnamefont {Brommel}},  \emph {et~al.},\ }\href {\doibase
  10.1103/PhysRevD.84.014503} {\bibfield  {journal} {\bibinfo  {journal}
  {Phys.Rev.}\ }\textbf {\bibinfo {volume} {D84}},\ \bibinfo {pages} {014503}
  (\bibinfo {year} {2011}{\natexlab{a}})},\ \Eprint
  {http://arxiv.org/abs/1012.4178} {arXiv:1012.4178 [hep-lat]} \BibitemShut
  {NoStop}%
%%CITATION = ARXIV:1012.4178;%%
\bibitem [{\citenamefont {Aoki}\ \emph
  {et~al.}(2011{\natexlab{b}})\citenamefont {Aoki} \emph
  {et~al.}}]{Aoki:2010dy}%
  \BibitemOpen
  \bibfield  {author} {\bibinfo {author} {\bibfnamefont {Y.}~\bibnamefont
  {Aoki}} \emph {et~al.} (\bibinfo {collaboration} {RBC Collaboration, UKQCD
  Collaboration}),\ }\href {\doibase 10.1103/PhysRevD.83.074508} {\bibfield
  {journal} {\bibinfo  {journal} {Phys.Rev.}\ }\textbf {\bibinfo {volume}
  {D83}},\ \bibinfo {pages} {074508} (\bibinfo {year} {2011}{\natexlab{b}})},\
  \Eprint {http://arxiv.org/abs/1011.0892} {arXiv:1011.0892 [hep-lat]}
  \BibitemShut {NoStop}%
\bibitem [{\citenamefont {Christ}\ \emph {et~al.}(2010)\citenamefont {Christ},
  \citenamefont {Dawson}, \citenamefont {Izubuchi}, \citenamefont {Jung},
  \citenamefont {Liu} \emph {et~al.}}]{Christ:2010dd}%
  \BibitemOpen
  \bibfield  {author} {\bibinfo {author} {\bibfnamefont {N.}~\bibnamefont
  {Christ}}, \bibinfo {author} {\bibfnamefont {C.}~\bibnamefont {Dawson}},
  \bibinfo {author} {\bibfnamefont {T.}~\bibnamefont {Izubuchi}}, \bibinfo
  {author} {\bibfnamefont {C.}~\bibnamefont {Jung}}, \bibinfo {author}
  {\bibfnamefont {Q.}~\bibnamefont {Liu}},  \emph {et~al.},\ }\href {\doibase
  10.1103/PhysRevLett.105.241601} {\bibfield  {journal} {\bibinfo  {journal}
  {Phys.Rev.Lett.}\ }\textbf {\bibinfo {volume} {105}},\ \bibinfo {pages}
  {241601} (\bibinfo {year} {2010})},\ \Eprint {http://arxiv.org/abs/1002.2999}
  {arXiv:1002.2999 [hep-lat]} \BibitemShut {NoStop}%
\bibitem [{\citenamefont {Ohki}\ \emph {et~al.}(2009)\citenamefont {Ohki} \emph
  {et~al.}}]{Ohki:2009mt}%
  \BibitemOpen
  \bibfield  {author} {\bibinfo {author} {\bibfnamefont {H.}~\bibnamefont
  {Ohki}} \emph {et~al.},\ }\href@noop {} {\bibfield  {journal} {\bibinfo
  {journal} {PoS}\ }\textbf {\bibinfo {volume} {LAT2009}},\ \bibinfo {pages}
  {124} (\bibinfo {year} {2009})},\ \Eprint {http://arxiv.org/abs/0910.3271}
  {arXiv:0910.3271 [hep-lat]} \BibitemShut {NoStop}%
%%CITATION = 0910.3271;%%
\bibitem [{\citenamefont {Aoki}\ \emph {et~al.}(2010)\citenamefont {Aoki} \emph
  {et~al.}}]{Aoki:2009ix}%
  \BibitemOpen
  \bibfield  {author} {\bibinfo {author} {\bibfnamefont {S.}~\bibnamefont
  {Aoki}} \emph {et~al.} (\bibinfo {collaboration} {PACS-CS}),\ }\href
  {\doibase 10.1103/PhysRevD.81.074503} {\bibfield  {journal} {\bibinfo
  {journal} {Phys. Rev.}\ }\textbf {\bibinfo {volume} {D81}},\ \bibinfo {pages}
  {074503} (\bibinfo {year} {2010})},\ \Eprint {http://arxiv.org/abs/0911.2561}
  {arXiv:0911.2561 [hep-lat]} \BibitemShut {NoStop}%
%%CITATION = 0911.2561;%%
\bibitem [{\citenamefont {Hasenfratz}\ \emph {et~al.}(2008)\citenamefont
  {Hasenfratz}, \citenamefont {Hoffmann},\ and\ \citenamefont
  {Schaefer}}]{Hasenfratz:2008fg}%
  \BibitemOpen
  \bibfield  {author} {\bibinfo {author} {\bibfnamefont {A.}~\bibnamefont
  {Hasenfratz}}, \bibinfo {author} {\bibfnamefont {R.}~\bibnamefont
  {Hoffmann}}, \ and\ \bibinfo {author} {\bibfnamefont {S.}~\bibnamefont
  {Schaefer}},\ }\href {\doibase 10.1103/PhysRevD.78.014515} {\bibfield
  {journal} {\bibinfo  {journal} {Phys.Rev.}\ }\textbf {\bibinfo {volume}
  {D78}},\ \bibinfo {pages} {014515} (\bibinfo {year} {2008})},\ \Eprint
  {http://arxiv.org/abs/0805.2369} {arXiv:0805.2369 [hep-lat]} \BibitemShut
  {NoStop}%
\bibitem [{\citenamefont {Luscher}\ and\ \citenamefont
  {Palombi}(2008)}]{Luscher:2008tw}%
  \BibitemOpen
  \bibfield  {author} {\bibinfo {author} {\bibfnamefont {M.}~\bibnamefont
  {Luscher}}\ and\ \bibinfo {author} {\bibfnamefont {F.}~\bibnamefont
  {Palombi}},\ }\href@noop {} {\bibfield  {journal} {\bibinfo  {journal} {PoS}\
  }\textbf {\bibinfo {volume} {LATTICE2008}},\ \bibinfo {pages} {049} (\bibinfo
  {year} {2008})},\ \Eprint {http://arxiv.org/abs/0810.0946} {arXiv:0810.0946
  [hep-lat]} \BibitemShut {NoStop}%
%%CITATION = 0810.0946;%%
\bibitem [{\citenamefont {DeGrand}(2008)}]{DeGrand:2008ps}%
  \BibitemOpen
  \bibfield  {author} {\bibinfo {author} {\bibfnamefont {T.}~\bibnamefont
  {DeGrand}},\ }\href {\doibase 10.1103/PhysRevD.78.117504} {\bibfield
  {journal} {\bibinfo  {journal} {Phys.Rev.}\ }\textbf {\bibinfo {volume}
  {D78}},\ \bibinfo {pages} {117504} (\bibinfo {year} {2008})},\ \Eprint
  {http://arxiv.org/abs/0810.0676} {arXiv:0810.0676 [hep-lat]} \BibitemShut
  {NoStop}%
\bibitem [{\citenamefont {Hasenbusch}(2001)}]{Hasenbusch:2001ne}%
  \BibitemOpen
  \bibfield  {author} {\bibinfo {author} {\bibfnamefont {M.}~\bibnamefont
  {Hasenbusch}},\ }\href@noop {} {\bibfield  {journal} {\bibinfo  {journal}
  {Phys. Lett.}\ }\textbf {\bibinfo {volume} {B519}},\ \bibinfo {pages} {177}
  (\bibinfo {year} {2001})},\ \Eprint {http://arxiv.org/abs/hep-lat/0107019}
  {hep-lat/0107019} \BibitemShut {NoStop}%
%%CITATION = HEP-LAT/0107019;%%
\bibitem [{\citenamefont {Allton}\ \emph {et~al.}(2008)\citenamefont {Allton}
  \emph {et~al.}}]{Allton:2008pn}%
  \BibitemOpen
  \bibfield  {author} {\bibinfo {author} {\bibfnamefont {C.}~\bibnamefont
  {Allton}} \emph {et~al.} (\bibinfo {collaboration} {RBC-UKQCD}),\ }\href
  {\doibase 10.1103/PhysRevD.78.114509} {\bibfield  {journal} {\bibinfo
  {journal} {Phys. Rev.}\ }\textbf {\bibinfo {volume} {D78}},\ \bibinfo {pages}
  {114509} (\bibinfo {year} {2008})},\ \Eprint {http://arxiv.org/abs/0804.0473}
  {arXiv:0804.0473 [hep-lat]} \BibitemShut {NoStop}%
%%CITATION = 0804.0473;%%
\bibitem [{\citenamefont {Luscher}(2010{\natexlab{a}})}]{Luscher:2010we}%
  \BibitemOpen
  \bibfield  {author} {\bibinfo {author} {\bibfnamefont {M.}~\bibnamefont
  {Luscher}},\ }\href@noop {} {\bibfield  {journal} {\bibinfo  {journal} {PoS}\
  }\textbf {\bibinfo {volume} {LATTICE2010}},\ \bibinfo {pages} {015} (\bibinfo
  {year} {2010}{\natexlab{a}})},\ \Eprint {http://arxiv.org/abs/1009.5877}
  {arXiv:1009.5877 [hep-lat]} \BibitemShut {NoStop}%
%%CITATION = 1009.5877;%%
\bibitem [{\citenamefont {Luscher}(2010{\natexlab{b}})}]{Luscher:2010iy}%
  \BibitemOpen
  \bibfield  {author} {\bibinfo {author} {\bibfnamefont {M.}~\bibnamefont
  {Luscher}},\ }\href {\doibase 10.1007/JHEP08(2010)071} {\bibfield  {journal}
  {\bibinfo  {journal} {JHEP}\ }\textbf {\bibinfo {volume} {08}},\ \bibinfo
  {pages} {071} (\bibinfo {year} {2010}{\natexlab{b}})},\ \Eprint
  {http://arxiv.org/abs/1006.4518} {arXiv:1006.4518 [hep-lat]} \BibitemShut
  {NoStop}%
%%CITATION = 1006.4518;%%
\bibitem [{\citenamefont {Morningstar}\ and\ \citenamefont
  {Peardon}(2004)}]{Morningstar:2003gk}%
  \BibitemOpen
  \bibfield  {author} {\bibinfo {author} {\bibfnamefont {C.}~\bibnamefont
  {Morningstar}}\ and\ \bibinfo {author} {\bibfnamefont {M.~J.}\ \bibnamefont
  {Peardon}},\ }\href {\doibase 10.1103/PhysRevD.69.054501} {\bibfield
  {journal} {\bibinfo  {journal} {Phys.Rev.}\ }\textbf {\bibinfo {volume}
  {D69}},\ \bibinfo {pages} {054501} (\bibinfo {year} {2004})},\ \Eprint
  {http://arxiv.org/abs/hep-lat/0311018} {arXiv:hep-lat/0311018 [hep-lat]}
  \BibitemShut {NoStop}%
%%CITATION = HEP-LAT/0311018;%%
\bibitem [{\citenamefont {de~Forcrand}\ \emph {et~al.}(1997)\citenamefont
  {de~Forcrand}, \citenamefont {Garcia~Perez},\ and\ \citenamefont
  {Stamatescu}}]{deForcrand:1997sq}%
  \BibitemOpen
  \bibfield  {author} {\bibinfo {author} {\bibfnamefont {P.}~\bibnamefont
  {de~Forcrand}}, \bibinfo {author} {\bibfnamefont {M.}~\bibnamefont
  {Garcia~Perez}}, \ and\ \bibinfo {author} {\bibfnamefont {I.-O.}\
  \bibnamefont {Stamatescu}},\ }\href {\doibase 10.1016/S0550-3213(97)00275-7}
  {\bibfield  {journal} {\bibinfo  {journal} {Nucl.Phys.}\ }\textbf {\bibinfo
  {volume} {B499}},\ \bibinfo {pages} {409} (\bibinfo {year} {1997})},\ \Eprint
  {http://arxiv.org/abs/hep-lat/9701012} {arXiv:hep-lat/9701012 [hep-lat]}
  \BibitemShut {NoStop}%
%%CITATION = HEP-LAT/9701012;%%
\bibitem [{\citenamefont {Allton}\ \emph {et~al.}(2007)\citenamefont {Allton}
  \emph {et~al.}}]{Allton:2007hx}%
  \BibitemOpen
  \bibfield  {author} {\bibinfo {author} {\bibfnamefont {C.}~\bibnamefont
  {Allton}} \emph {et~al.} (\bibinfo {collaboration} {RBC and UKQCD
  Collaborations}),\ }\href {\doibase 10.1103/PhysRevD.76.014504} {\bibfield
  {journal} {\bibinfo  {journal} {Phys.Rev.}\ }\textbf {\bibinfo {volume}
  {D76}},\ \bibinfo {pages} {014504} (\bibinfo {year} {2007})},\ \Eprint
  {http://arxiv.org/abs/hep-lat/0701013} {arXiv:hep-lat/0701013 [hep-lat]}
  \BibitemShut {NoStop}%
%%CITATION = HEP-LAT/0701013;%%
\bibitem [{\citenamefont {Bardeen}\ \emph {et~al.}(2001)\citenamefont
  {Bardeen}, \citenamefont {Duncan}, \citenamefont {Eichten}, \citenamefont
  {Isgur},\ and\ \citenamefont {Thacker}}]{Bardeen:2001jm}%
  \BibitemOpen
  \bibfield  {author} {\bibinfo {author} {\bibfnamefont {W.~A.}\ \bibnamefont
  {Bardeen}}, \bibinfo {author} {\bibfnamefont {A.}~\bibnamefont {Duncan}},
  \bibinfo {author} {\bibfnamefont {E.}~\bibnamefont {Eichten}}, \bibinfo
  {author} {\bibfnamefont {N.}~\bibnamefont {Isgur}}, \ and\ \bibinfo {author}
  {\bibfnamefont {H.}~\bibnamefont {Thacker}},\ }\href {\doibase
  10.1103/PhysRevD.65.014509} {\bibfield  {journal} {\bibinfo  {journal}
  {Phys.Rev.}\ }\textbf {\bibinfo {volume} {D65}},\ \bibinfo {pages} {014509}
  (\bibinfo {year} {2001})},\ \Eprint {http://arxiv.org/abs/hep-lat/0106008}
  {arXiv:hep-lat/0106008 [hep-lat]} \BibitemShut {NoStop}%
%%CITATION = HEP-LAT/0106008;%%
\end{thebibliography}%

\end{document}